\documentclass[intlimits,twoside,a4paper]{article}

\usepackage{amsmath,amssymb}
\usepackage{graphicx}

\usepackage[T2A]{fontenc}
\usepackage[cp1251]{inputenc}

\usepackage[eqsecnum]{cmpj3}

\issue{2018}{21}{2}{23801}
\doinumber{10.5488/CMP.21.23801}

\title[Spatiotemporal pattern formation in a three-variable CO oxidation reaction model]%
{Spatiotemporal pattern formation in a three-variable CO oxidation reaction model}
\author[I.S. Bzovska, I.M. Mryglod]{I.S. Bzovska, I.M. Mryglod}

\address{Institute for Condensed
Matter Physics of the National Academy of Sciences of Ukraine,\\
1~Svientsitskii~St., 79011 Lviv, Ukraine}

\date{Received February 16, 2018, in final form May 2, 2018}

\begin{document}

\maketitle

\begin{abstract}
The spatiotemporal pattern formation is studied in the catalytic carbon
monoxide oxidation reaction that takes into account the diffusion
processes over the Pt(110) surface, which may contain structurally
different areas. These areas are formed during
CO-induced transition from a reconstructed phase with $1\times2$
geometry of the overlayer to a bulk-like ($1\times1$) phase with
square atomic arrangement. Despite the CO oxidation reaction
being non-autocatalytic, we have shown that the analytic conditions
of the existence of the Turing and the Hopf bifurcations can be
satisfied in such systems. Thus, the system may lose its stability
in two ways --- either through the Hopf bifurcation leading to the
formation of temporal patterns in the system or through the Turing
bifurcation leading to the formation of regular spatial patterns.
At a simultaneous implementation of both scenarios, spatiotemporal
patterns for CO and oxygen coverages are obtained in the system.
\keywords reaction-diffusion model, spatiotemporal patterns, the
Hopf bifurcation, the Turing bifurcation
\pacs 82.40.Bj, 82.45.Jn
\end{abstract}

\section{Introduction}
\label{Intro}

In recent times, spatiotemporal pattern formation in spatially
extended systems, such as reaction-diffusion systems, has been
extensively studied \cite{Ebeling,Cross,I1,Khar}. In these systems,
the concentration of one or more substances distributed in space
can change under the influence of two processes: local chemical
reactions in which the substances transform into each other,
and diffusion which causes the substances to spread out over a
surface in space.

Among chemical systems, the catalytic oxidation of CO on platinum
(110) is one of the most prominent examples of a
reaction-diffusion system showing a variety of complex
spatiotemporal patterns \cite{14,3,13,pavl}. For this system,
various experiments on pattern formation have been carried out.
Pattern formation was monitored by means of photoemission electron
microscopy (PEEM) \cite{exp4,exp3,exp5}. The experimental
parameters were chosen such that the reaction was oscillatory and,
furthermore, uniform oscillations were unstable and a complex
state of spiral-wave turbulence spontaneously developed.

An orientation of the catalyst surface in such systems 
decisively influences  the occurrence of oscillations and surface
patterns \cite{I1,14,3}. A clean Pt(110) top surface layer
reconstructs into a $1\times2$ ``missing row'' structure. This
reconstruction can be reversibly lifted by adsorption of CO
molecules. Since oxygen adsorption is favoured on the
unreconstructed $1\times1$ phase, a periodic switching can occur between two
states of a different catalytic activity resulting in
temporal oscillations of the reaction rate. Local spatial coupling
across the catalytic surface is provided by surface diffusion of
adsorbed CO and oxygen. Under such oscillatory conditions, the
interplay between the reaction and diffusion processes can lead to the
development of spatiotemporal patterns.

The formation of spatiotemporal patterns occurs under two main
symmetry-breaking instabilities such as the Hopf and the Turing ones
\cite{1,4}. An interaction and competition of these bifurcations
have been considered for different reaction-diffusion systems,
including Belousov-Zhabotinsky autocatalytic reaction
\cite{vanag}, the FitzHugh-Nagumo model \cite{1,4}, etc. In these
models, a variety of modes has been received, including mixed modes, i.e.,
 spatial patterns modulated in time.

In this paper, we study the mechanisms of spatiotemporal pattern
formation in the carbon monoxide oxidation reaction on the surface
of Pt(110). A simple three-variable model has been developed to
account for most of the dynamic features of the reaction. The main
purpose of the study is to determine whether  a combined scenario of the formation of dissipative
patterns is possible in this three-variable model and under what
conditions. The analysis of
instabilities in time and space of the system is based on the methods
of linear stability theory and numerical modelling. It is shown
that conditions for the existence of the Turing and of the Hopf
bifurcations can be satisfied in such non-autocatalytic systems.
As a result, this leads to the formation of dissipative patterns.

The paper is organized as follows. A model of the catalytic
oxidation reaction of carbon monoxide and the linear stability
theory are introduced in the next section. In section
\ref{Results}, the results of our calculations and a discussion of
the obtained results are presented.  The paper ends
with conclusions in section \ref{Concl}.

\section{Model and theory}
\label{Model}

Let us consider a model of the catalytic oxidation reaction of
carbon monoxide that takes the diffusion processes over the
Pt(110) surface into account. The model was first introduced in
1992 by Krischer, Eiswirth and Ertl \cite{Krischer} without the
diffusion terms. Later on, it was extended to include diffusion terms
for the CO species by B\"ar et al. to account for the pattern
formation behaviours observed in experiments \cite{Bar}. We include
the diffusion terms into the system of kinetic
differential equations for all species \cite{2,prep2015}:
\begin{align}
\label{1.1} &\displaystyle\frac{{\rm d}\theta_{\rm CO}}{{\rm
d}\tau}=D_{1}\Delta\theta_{\rm CO}+p_{\rm CO}k_{\rm CO}s_{\rm CO}(1-\theta_{\rm CO}^{q})-d\theta_{\rm CO}-k_{\text r}\theta_{\rm CO}\theta_{\rm O}\,, \\
\label{1.2} &\displaystyle\frac{{\rm d}\theta_{\rm O}}{{\rm
d}\tau}=D_{2}\Delta\theta_{\rm O}+p_{\rm O_{2}}k_{\rm
O}[s_{1\times1}\theta_{1\times1}+s_{1\times2}(1-\theta_{1\times1})](1-\theta_{\rm
CO}-\theta_{\rm O})^{2}-k_{\text r}\theta_{\rm
CO}\theta_{\rm O}\,, \\
\label{1.3} &\displaystyle\frac{{\rm d}\theta_{1\times1}}{{\rm
d}\tau}=D_{3}\Delta\theta_{\rm
1\times1}+k_{5}\left\{\left[1+\exp\displaystyle\left(\frac{u_{0}-\theta_{\rm
CO}}{\delta u}\right)\right]^{-1}-\theta_{1\times1}\right\}.
\end{align}
Equation $(\ref{1.1})$ describes the change of the number of
adsorbed CO that takes into account the chemical reaction with
adsorbed oxygen, desorption of CO with desorption constant $d$ and
diffusion of CO. Equation $(\ref{1.2})$ describes the diffusion of
oxygen, its dissociative adsorption and changes due to CO
oxidation reaction. In many models, the diffusion of oxygen is
usually neglected compared to that of CO. Here, we take it into
account and consider the diffusion coefficient of adsorbed oxygen
being about three orders of magnitude lower than the CO diffusion
parameter \cite{pavl}. Equation $(\ref{1.3})$ is a kinetic
equation for the surface transformation. Function
$\{1+\exp\displaystyle[(u_{0}-\theta_{\rm CO})/\delta u]\}^{-1}$
is a nondecreasing and smooth function of $\theta_{\rm CO}$ at the
interval [0,1], which allows us to describe the transformation of
the reconstructed $1\times2$ surface structure into the $1\times1$
structure depending on the amount of CO coverage \cite{3}. For an
inhomogeneous surface, the Laplacian term $\Delta\theta_{\rm
1\times1}$ in equation~$(\ref{1.3})$ originates from the
contribution of the interfaces between different surface
geometries to the total system energy \cite{pavl}. Consequently,
the coefficient $D_{3}$ describes the energy costs of such
interfaces. In this model, the precursor-type kinetics of CO
adsorption is accounted for by the exponent $q=3$ in the right-hand side of equation ($\ref{1.1}$). It makes the model more realistic
since the inhibition of adsorption of CO and O$_{2}$ is asymmetric,
and the preadsorbed CO blocks the oxygen adsorption but not vice versa.
A more detailed explanation and values of the parameters used in further
calculations are presented in table~\ref{tab1}.

\begin{table}[!t]
 \caption{Parameters of the model  \cite{3}.} \centering
\vspace{2ex}
\renewcommand{\arraystretch}{1.15} 
\begin{tabular}{c|c|c}
\hline\hline
$T$ & 540 K&Temperature\\
 \hline
  $p_{\rm O_{2}}$ & $9.75\times10^{-5}$ Torr& O$_{2}$ partial pressure\\
 \hline
  $k_{\rm CO}$ & $4.2\times10^{5}$ s$^{-1}$Torr$^{-1}$&Impingement rate of CO\\
\hline
 $k_{\rm O}$ & $7.8\times10^{5}$ s$^{-1}$Torr$^{-1}$&Impingement rate of O$_{2}$\\
\hline
 $d$ & 10.21 s$^{-1}$&CO desorption rate\\
 \hline
 $D_{1}$ & 10$^{-7}$ cm$^{2}$s$^{-1}$&CO diffusion rate\\
 \hline
 $D_{2}$ & 10$^{-10}$ cm$^{2}$s$^{-1}$&O diffusion rate\\
 \hline
 $k_{\text r}$ & 283.8 s$^{-1}$&Reaction rate\\
\hline
  $s_{\rm CO}$ & 1&CO sticking coefficient\\
\hline $s_{\rm O,1\times2}$& 0.4&Oxygen sticking coefficient on
the
$1\times2$ phase\\
\hline
$u_{0}$, $\delta u$ & 0.35, 0.05&Parameters for the structural phase transition\\
\hline
$k_{5}$ &1.61 s$^{-1}$&Phase transition rate\\
\hline \hline
\end{tabular}
\label{tab1}
\end{table}

System (\ref{1.1})--(\ref{1.3}) can be transformed by substitution
\begin{eqnarray}
&&t=k_{\text r}\tau, \qquad \bar{D}_{1,2,3}=D_{1,2,3}/k_{\text r}\,, \qquad
\bar{p}_{\rm CO}=p_{\rm CO}k_{\rm CO}s_{\rm
CO}/k_{\text r}\,,\nonumber\\
&&\bar{p}_{\rm O_{2}}=p_{\rm O_{2}}k_{\rm O}s_{\rm
O}^{1\times2}/k_{\text r}\,, \qquad  \bar{d}=d/k_{\text r}\,, \qquad
\bar{k}_{5}=k_{5}/k_{\text r}\nonumber
\end{eqnarray}
into the following dimensionless form:
\begin{align}
\label{1a} &\displaystyle\frac{{\rm d}\theta_{\rm CO}}{{\rm
d}t}=F_{1}(\theta_{\rm CO},\theta_{\rm
O})=\bar{D}_{1}\Delta\theta_{\rm CO}+\bar{p}_{\rm CO}
(1-\theta_{\rm CO}^{3})-\bar{d}\theta_{\rm CO}-\theta_{\rm CO}\theta_{\rm O}\,, 
\end{align}
\begin{align}
\label{1b} &\displaystyle\frac{{\rm d}\theta_{\rm O}}{{\rm
d}t}=F_{2}(\theta_{\rm CO},\theta_{\rm
O},\theta_{1\times1})=\bar{D}_{2}\Delta\theta_{\rm O}+\bar{p}_{\rm
O_{2}}(1+\theta_{1\times1})(1-\theta_{\rm
CO}-\theta_{\rm O})^{2}-\theta_{\rm CO}\theta_{\rm O}\,, \\
\label{1c} &\displaystyle\frac{{\rm d}\theta_{1\times1}}{{\rm
d}t}=F_{3}(\theta_{\rm
CO},\theta_{1\times1})=\bar{D}_{3}\Delta\theta_{1\times1}+\bar{k}_{5}\left\{\left[1+\exp\displaystyle\left(\frac{u_{0}-\theta_{\rm
CO}}{\delta u}\right)\right]^{-1}-\theta_{1\times1}\right\}.
\end{align}
\looseness=-1 $s_{\rm O}=s_{\rm O}^{1\times1}\theta_{1\times1}+s_{\rm
O}^{1\times2}(1-\theta_{1\times1})=s_{\rm
O}^{1\times2}(1+\theta_{1\times1})$ under the assumption that for
Pt(110) we have $s_{\rm O}^{1\times1}/s_{\rm O}^{1\times2}\simeq
2$.

The system of differential equations $(\ref{1a})$--$(\ref{1c})$
with partial derivatives cannot be solved analytically.
Therefore, the analysis of the system instabilities in time and space
has been based on the methods of the linear stability theory and on
numerical simulations. The system of equations
$(\ref{1a})$--$(\ref{1c})$ in the linear approximation for the
deviations from steady state
$\delta\theta_{i}(\mathbf{r},t)=\theta_{i}(\mathbf{r},t)-\theta_{i,\text{s}}(\mathbf{r})$
looks as follows:
\begin{eqnarray}
\frac{\partial}{\partial
t}\delta\theta_{i}(\mathbf{r},t)=\sum_{j=1}^{3}\left(\frac{\partial
F_{i}}{\partial
\theta_{j}}\right)_{\theta_{k}=\theta_{k,\text{s}}}\delta\theta_{j}(\mathbf{r},t)+\bar{D}_{i}\Delta
\delta\theta_{i}(\mathbf{r},t), \qquad i,j=\rm CO,O,1\times1.
\nonumber
\end{eqnarray}

Stability of the system has been investigated using the method of
normal modes concerning the perturbation periodic in space (normal
mode) with a wavelength $\lambda$. To this end, we do the following
substitution $\delta\theta_{i}(\mathbf{r},t)\sim \re ^{\omega
t+\text{i}\mathbf{kr}}$, where $k=|\mathbf{k}|=2\piup/\lambda$ is a wave
number, and we obtain the following linear system of equations
\begin{eqnarray}
\sum_{j=1}^{3}\left[\left(\frac{\partial F_{i}}{\partial
\theta_{j}}\right)_{\theta_{k}=\theta_{k,\text{s}}}-\bar{D}_{i}k^{2}\delta_{ij}-\omega\delta_{ij}\right]\delta\theta_{j}=0,
\qquad i=1, 2, 3.
\end{eqnarray}
Stability analysis requires a solution of the secular equation
\begin{eqnarray}
\label{sec} \det\left|\left|\left(\displaystyle\frac{\partial
F_{i}}{\partial
\theta_{j}}\right)_{\theta_{k}=\theta_{k,\text{s}}}-\bar{D}_{i}k^{2}\delta_{ij}-\omega\delta_{ij}\right|\right|=0,
\end{eqnarray}
whereof we get an equation for $\omega(k)$:
\begin{eqnarray}
\label{2} \omega^{3}-b(k)\omega^{2}+c(k)\omega-d(k)=0,
\end{eqnarray}
where we have introduced the next notations:
\begin{align}
b(k)&=\sigma-k^{2}\left(\bar{D}_{1}+\bar{D}_{2}+\bar{D}_{3}\right),\nonumber\\
c(k)&=\Sigma-k^{2}\left[\bar{D}_{1}(a_{22}+a_{33})+
\bar{D}_{2}(a_{11}+a_{33})+\bar{D}_{3}(a_{11}+a_{22})\right]+k^{4}\left(\bar{D}_{1}\bar{D}_{2}+\bar{D}_{1}\bar{D}_{3}+\bar{D}_{2}\bar{D}_{3}\right),\nonumber\\
d(k)&=\Delta-k^{2}\sum_{i=1}^{3}\bar{D}_{i}\eta_{i}+
k^{4}\left(a_{11}\bar{D}_{2}\bar{D}_{3}+a_{22}\bar{D}_{1}\bar{D}_{3}
+a_{33}\bar{D}_{1}\bar{D}_{2}\right)
-k^{6}\bar{D}_{1}\bar{D}_{2}\bar{D}_{3}.\nonumber
\end{align}
Here, $a_{ij}=(\partial F_{i}/\partial
\theta_{j})_{\theta_{k}=\theta_{k,\text{s}}}$,
$\sigma=a_{11}+a_{22}+a_{33}$ is the trace of the characteristic
matrix $\{a_{ij}\}$,
$\Delta=a_{11}(a_{22}a_{33}-a_{23}a_{32})-a_{12}(a_{21}a_{33}-a_{23}a_{31})+a_{13}(a_{21}a_{32}-a_{22}a_{31})$
is its determinant, $\Sigma=\sum_{i=1}^{3}\eta_{i}$, where
$\eta_{i}=a_{jj}a_{ll}-a_{jl}a_{lj}$, $i\neq j\neq l$. For our
model,
\begin{align}
a_{11}&=-3\bar{p}_{\rm CO}\theta_{{\rm
CO},\text{s}}^{2}-\bar{d}-\theta_{{\rm O},\text{s}}\,, \qquad a_{12}=-\theta_{{\rm CO},\text{s}}\,, \qquad a_{13}=0,\nonumber\\
a_{21}&=-2\bar{p}_{\rm
O_{2}}(1+\theta_{1\times1,\text{s}})(1-\theta_{{\rm CO},\text{s}}-\theta_{{\rm O},\text{s}})-\theta_{{\rm O},\text{s}}\,, \nonumber\\
a_{22}&=-2\bar{p}_{\rm
O_{2}}(1+\theta_{1\times1,\text{s}})(1-\theta_{{\rm CO},\text{s}}-\theta_{{\rm O},\text{s}})-\theta_{{\rm CO},\text{s}}\,,\nonumber\\
a_{23}&=\bar{p}_{\rm
O_{2}}(1-\theta_{{\rm CO},\text{s}}-\theta_{{\rm O},\text{s}})^{2},\nonumber\\
a_{31}&=\frac{\bar{k}_{5}}{\delta
u}\frac{\exp\left(\frac{u_{0}-\theta_{{\rm
CO},\text{s}}}{\delta
u}\right)}{\left[1+\exp\left(\frac{u_{0}-\theta_{{\rm
CO},\text{s}}}{\delta u}\right)\right]^{2}}\,, \qquad a_{32}=0, \qquad
a_{33}=-\bar{k}_{5}. \label{char}
\end{align}

Equation ($\ref{2}$) is a cubic equation with real coefficients.
In a general case, its solutions can contain both real and imaginary
parts, i.e., $\omega(k)=\Re\omega(k)+\text{i}\Im\omega(k)$. The
component $\Re\omega(k)$ describes the stability of a solution
$\big(\delta\theta_{{\rm CO},k}(\omega)$, $\delta\theta_{{\rm
O},k}(\omega)$, $\delta\theta_{{\rm 1\times1},k}(\omega)\big)$ and
defines the process of relaxation, while $\Im\omega(k)$ sets the
frequency of the oscillating process. The system is stable if
\begin{eqnarray}
\label{3}  \Re\omega(k)<0  \qquad {\rm for} \; \forall k,
\end{eqnarray}
that is, when all normal modes are exponentially reduced. In the
case when at least for one mode at a certain $k$ inequality
$\Re\omega(k)>0$ becomes true, the whole system becomes unstable
because the amplitude of the corresponding motion increases.

It is well-known that coefficients of a cubic equation and its
roots are connected by the relations:
\begin{align}
\label{omega}
b&=\omega_{1}+\omega_{2}+\omega_{3}\,,\nonumber\\
c&=\omega_{1}\omega_{2}+\omega_{1}\omega_{3}+\omega_{2}\omega_{3}\,,\nonumber\\
d&=\omega_{1}\omega_{2}\omega_{3}\,,\nonumber\\
bc-d&=(\omega_{1}+\omega_{2})(\omega_{1}+\omega_{3})(\omega_{2}+\omega_{3}).
\end{align}
Consequently, as follows from relations ($\ref{3}$) and
($\ref{omega}$), a homogeneous state of the whole system is
stable if
\begin{eqnarray}
\label{ineq} b<0, \qquad c>0,\qquad d<0,\qquad bc-d<0.
\end{eqnarray}
The violation of any of inequalities ($\ref{ineq}$) means that  a bifurcation has occurred in
the system. The broken condition $d<0$
means the appearance of one real positive eigenvalue in the
system. The broken condition $bc-d<0$ means that there are two complex
conjugate eigenvalues with a positive real part. The first case
corresponds to the Turing bifurcation, and the second corresponds to the
Hopf one.

Function $d(k^{2})=\Delta-\alpha_{\text T}k^{2}+\beta_{\text
T}k^{4}-\delta_{\text T}k^{6}$ is a cubic parabola which has local
extremes. The maximum is
\begin{eqnarray}
d_{\rm max}(k_{\text
T}^{2})=\Delta+\displaystyle\frac{1}{27\delta_{\text T}^{2}}
\left[2(\beta_{\text T}^{2}-3\alpha_{\text T}\delta_{\text
T})^{\frac{3}{2}}+\beta_{\text T}(2\beta_{\text
T}^{2}-9\alpha_{\text T}\delta_{\text T})\right]
\end{eqnarray}
and is reached at the point $k_{\text T}^{2}=(\beta_{\text
T}+\sqrt{\beta_{\text T}^{2}-3\alpha_{\text T}\delta_{\text
T}})/3\delta_{\text T}^{2}$. Here, $\alpha_{\text
T}=\sum_{i=1}^{3}\bar{D}_{i}\eta_{i}$, $\beta_{\text
T}=a_{11}\bar{D}_{2}\bar{D}_{3}+a_{22}
\bar{D}_{1}\bar{D}_{3}+a_{33}\bar{D}_{1}\bar{D}_{2}$,
$\delta_{\text T}=\bar{D}_{1}\bar{D}_{2}\bar{D}_{3}$. For the
Turing bifurcation, it is necessary that in a certain range of wave
numbers, $d(k^{2})$ should become greater than zero,
\begin{eqnarray}
d_{\rm max}(k_{\text T}^{2})>0. \label{cond}
\end{eqnarray}
In \cite{Borina}, it is affirmed that this is possible only if at
least one of the coefficients on the main diagonal of matrix
$\{a_{ij}\}$ is greater than zero (a well-known condition for the
existence of autocatalysis).

The condition for the Hopf bifurcation can be obtained in a
similar way and is as follows:
\begin{eqnarray}
\label{cond2} F_{\rm max}(k_{0}^{2})=bc-d=\sigma\Sigma-\Delta
+\displaystyle\frac{1}{27\delta_{\text V}^{2}}
\left[2(\beta_{\text V}^{2}-3\alpha_{\text V}\delta_{\text
V})^{\frac{3}{2}}+\beta_{\text V}(2\beta_{\text
V}^{2}-9\alpha_{\text V}\delta_{\text V})\right]>0,
\end{eqnarray}
where $k_{0}^{2}=(\beta_{\text V}+\sqrt{\beta_{\text
V}^{2}-3\alpha_{\text V}\delta_{\text V}})/3\delta_{\text V}^{2}$.
Here,
\begin{align}
\alpha_{\text
V}&=\bar{D}_{1}(\sigma^{2}-a_{11}^{2}-a_{13}a_{31}-a_{12}a_{21})
+\bar{D}_{2}(\sigma^{2}-a_{22}^{2}-a_{23}a_{32}-a_{12}a_{21})\nonumber\\
&\quad+\bar{D}_{3}(\sigma^{2}-a_{33}^{2}-a_{13}a_{31}-a_{23}a_{32}),\nonumber\\
\beta_{\text
V}&=(\bar{D}_{1}+\bar{D}_{3})(\bar{D}_{2}+\bar{D}_{3})(a_{11}+a_{22})
+(\bar{D}_{1}+\bar{D}_{2})(\bar{D}_{2}+\bar{D}_{3})(a_{11}+a_{33})\nonumber\\
&\quad+(\bar{D}_{1}+\bar{D}_{2})(\bar{D}_{1}+\bar{D}_{3})(a_{22}+a_{33}),\nonumber\\
\delta_{\text
V}&=(\bar{D}_{1}+\bar{D}_{2})(\bar{D}_{1}+\bar{D}_{3})(\bar{D}_{2}+\bar{D}_{3}).\nonumber
\end{align}
Again, to satisfy the inequality $(\ref{cond2})$, the sum of two
coefficients on the main diagonal of matrix $\{a_{ij}\}$ must be
greater than zero \cite{Borina}.

As we see, our diagonal coefficients $a_{11}$, $a_{22}$ and
$a_{33}$ are negative for all values of the system parameters.
This means that the catalytic CO oxidation reaction is not
autocatalytic because, as it was mentioned above, for autocatalytic
reactions, at least one of the diagonal coefficients must be
greater than zero. Nevertheless, we further show that conditions
of the existence of the Turing $(\ref{cond})$ and the Hopf
$(\ref{cond2})$ bifurcations can be satisfied in our
non-autocatalytic system at certain values of the system
parameters. We associate the emergence of these instabilities with
an interaction of nonlinear local transformations with a positive
feedback (i.e., surface phase transitions) and transport processes
(diffusion) which spatially couple the system.

\section{Results and discussion}
\label{Results}
\subsection{Parametric analysis}

\begin{figure}[!t]
\centering
   \includegraphics[clip=true,keepaspectratio=true,width=0.5\textwidth]{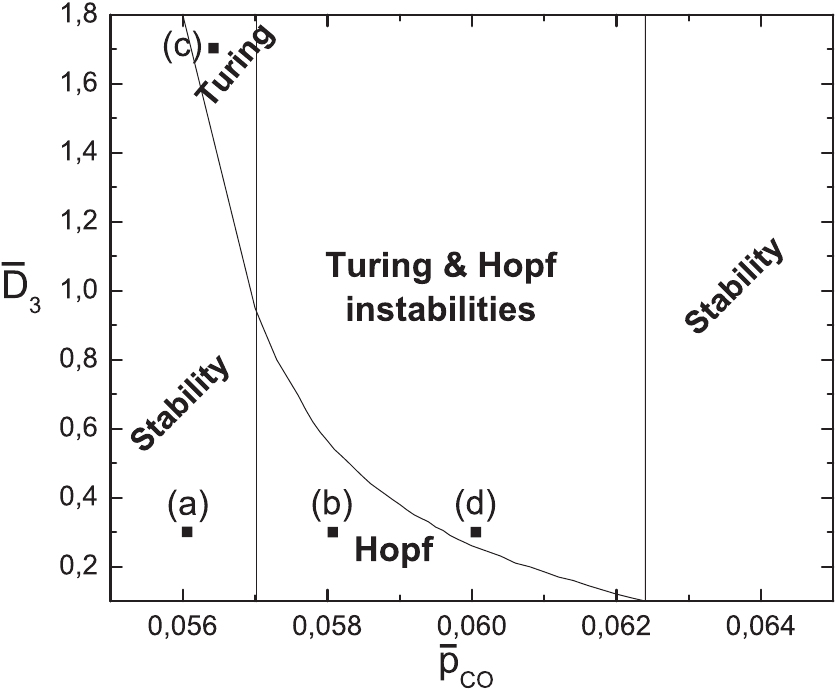}
   \caption{\label{StabDiag} Stability diagram of the model in ($\bar{p}_{\rm CO}, \; \bar{D}_{3}$) parameter space.}
  \end{figure}

Using conditions of the existence of the Turing $(\ref{cond})$ and
the Hopf $(\ref{cond2})$ bifurcations, we choose the regions in
($\bar{p}_{\rm CO}$, $\bar{D}_{3}$) parameter space which
correspond to these instabilities (see figure~$\ref{StabDiag}$).
The region of their intersection corresponds to the presence of
both instabilities. In the regions of low and high partial
pressures $\bar{p}_{\text{CO}}$, the system is stable. The region of intermediate partial
pressures $\bar{p}_{\text{CO}}$ where the oscillations occur coincides with the region of the 
structural surface transformation which is in agreement with the experimental observations by LEED \cite{cit8}. The values of the parameters used in the
calculations are presented in table~\ref{tab1}.

\begin{figure}[!t]
  \centering
  a)\raisebox{-.5\height}{\includegraphics[clip=true,keepaspectratio=true,width=0.35\textwidth]{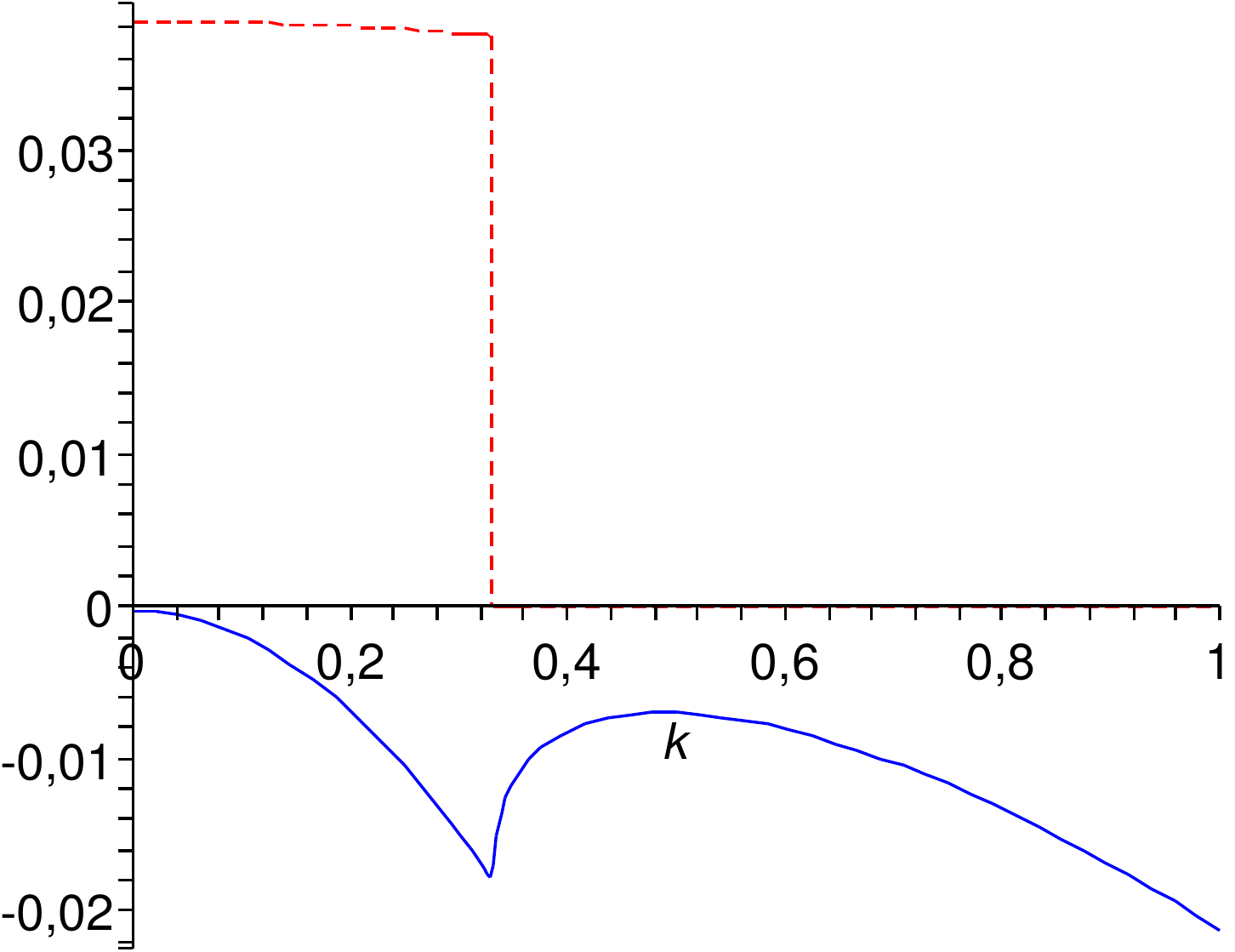}} \qquad
  b)\raisebox{-.5\height}{\includegraphics[clip=true,keepaspectratio=true,width=0.35\textwidth]{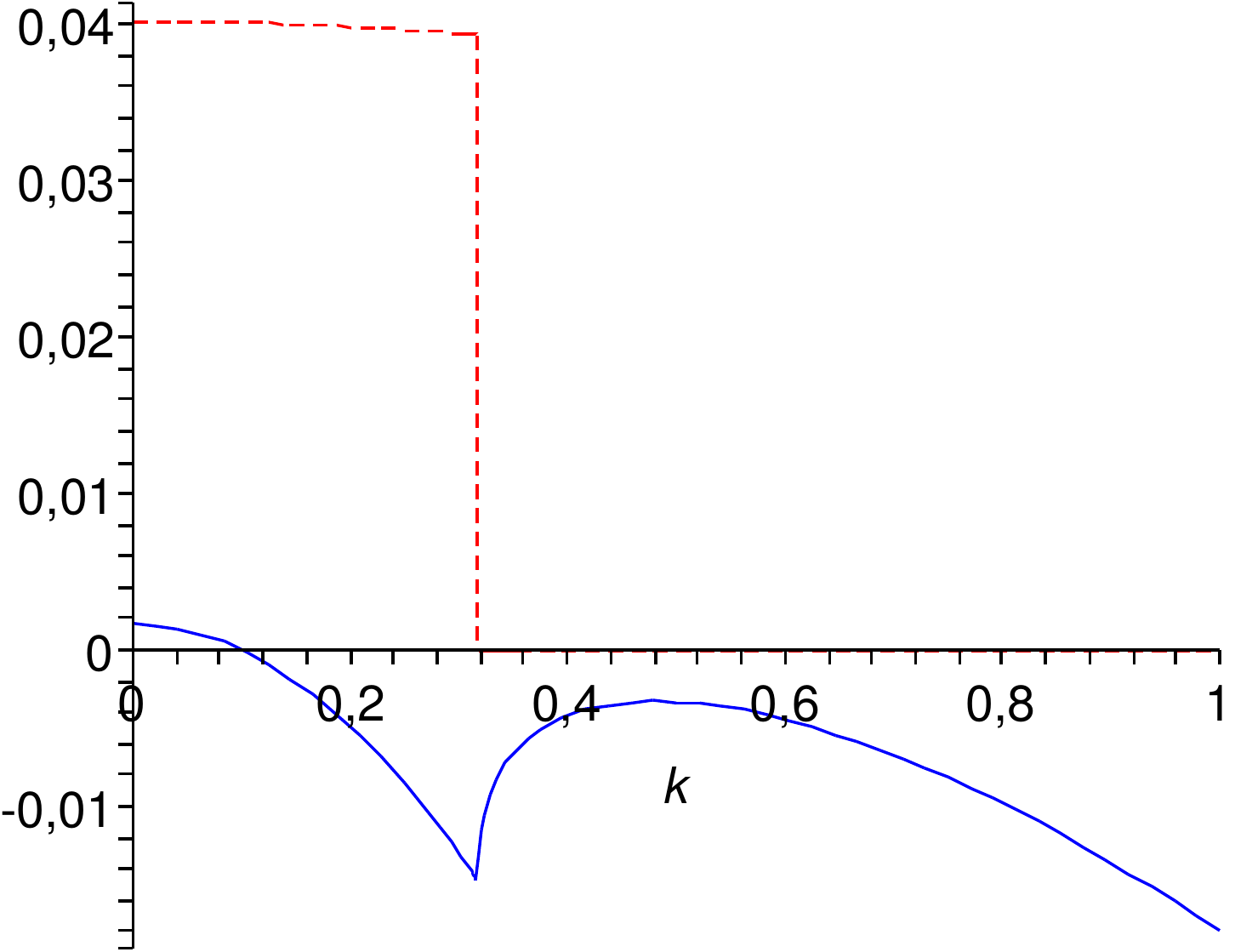}}\vspace{2mm}\\
  c)\raisebox{-.5\height}{\includegraphics[clip=true,keepaspectratio=true,width=0.35\textwidth]{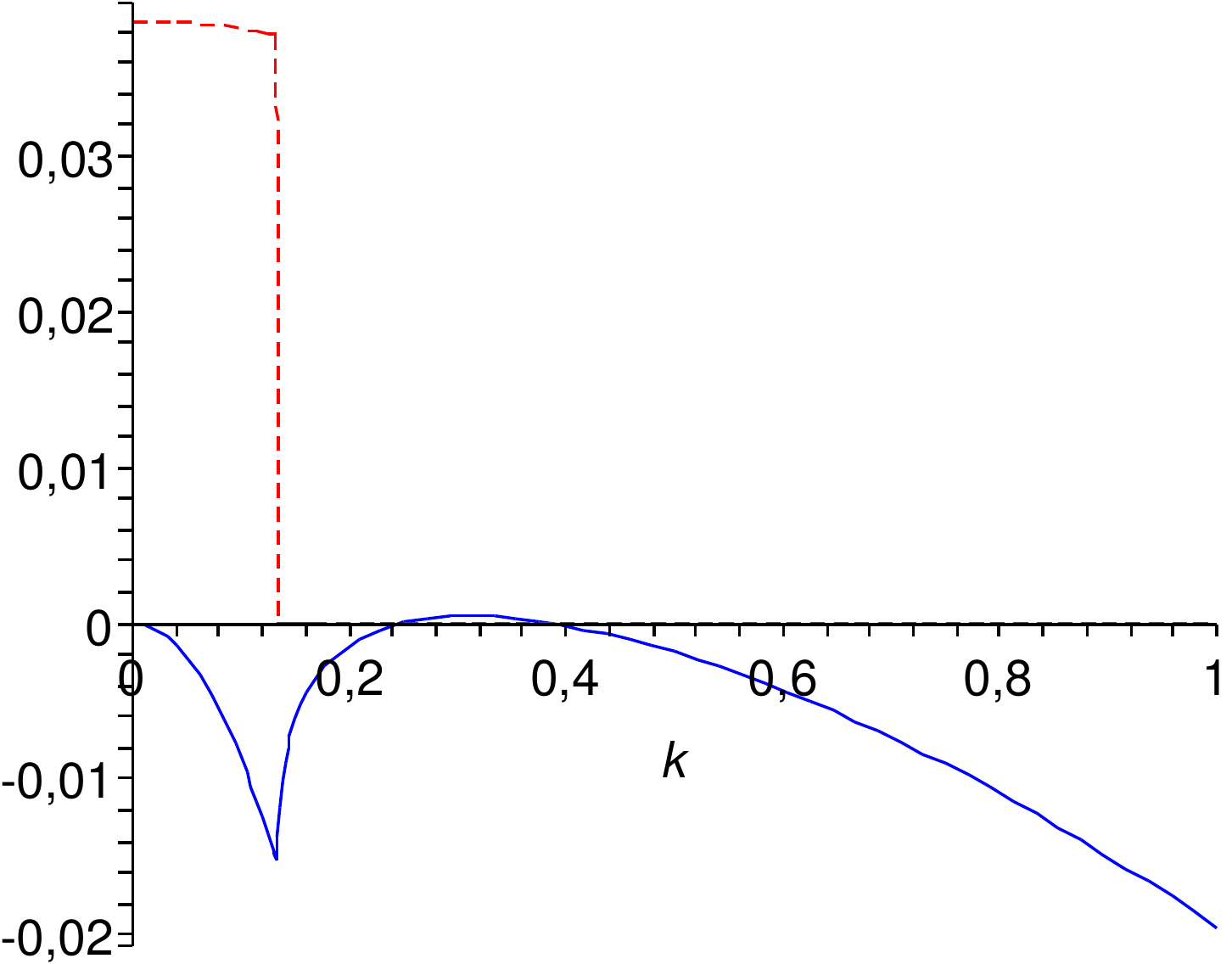}} \qquad
  d)\raisebox{-.5\height}{\includegraphics[clip=true,keepaspectratio=true,width=0.35\textwidth]{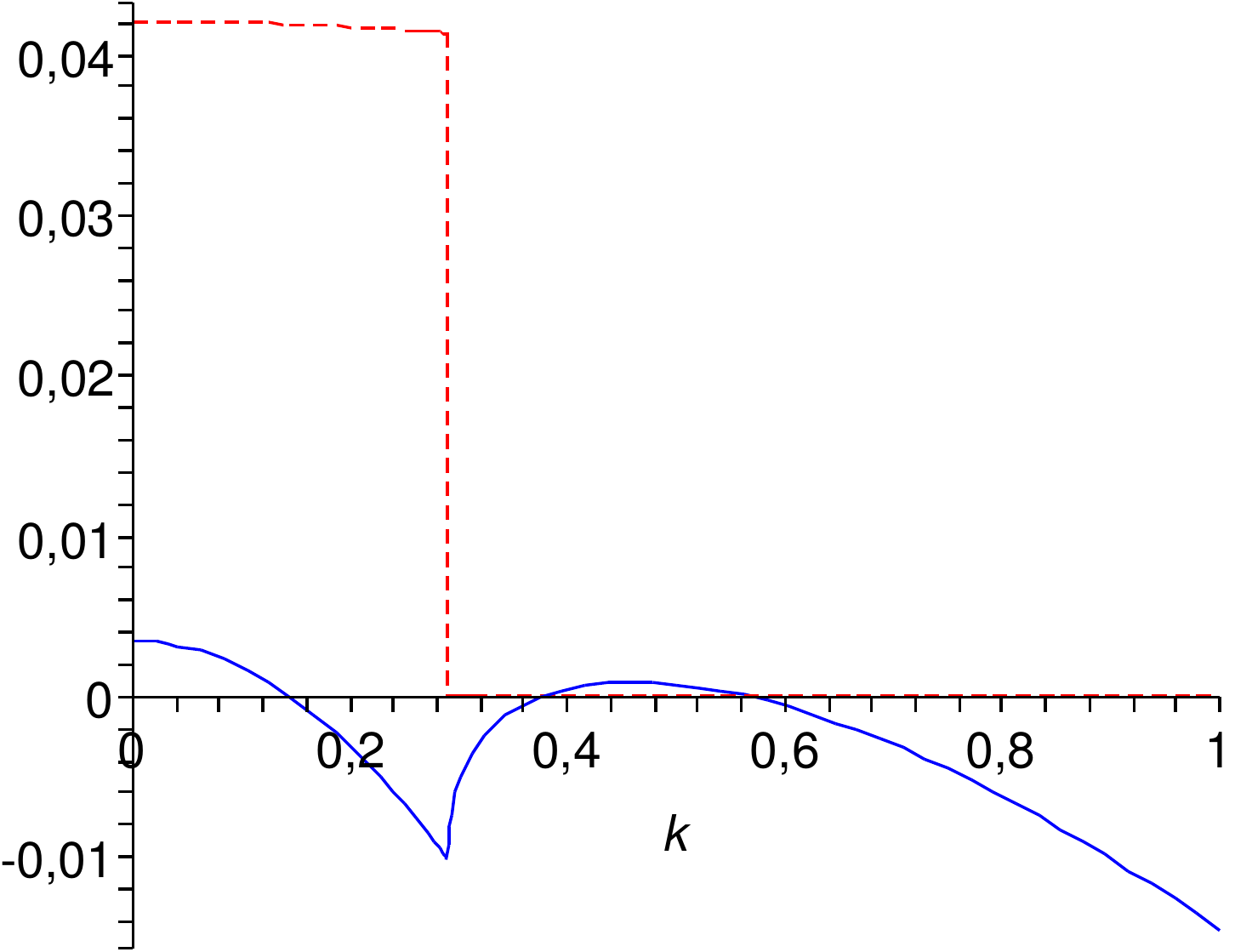}}
  \caption{(Colour online) Dispersion dependences of real $\Re\omega$ (solid line)
    and imaginary $\Im\omega/10$ (dashed line) parts
    on the wave number $k$ for the following sets of model parameters: (a)~$\bar{p}_{\rm CO}=0.056$,
    $\bar{D}_{3}=0.3$ (stable region); (b)~$\bar{p}_{\rm CO}=0.058$,
    $\bar{D}_{3}=0.3$ (Hopf instability);  (c)~$\bar{p}_{\rm CO}=0.0563$, $\bar{D}_{3}=1.7$ (Turing instability); and
    (d)~$\bar{p}_{\rm CO}=0.06$, $\bar{D}_{3}=0.3$ (both instabilities), respectively.
    Other model parameters are the same in all cases.} \label{fig1}
\end{figure}

Figure~$\ref{fig1}$ presents the dispersion dependences of real
$\Re\omega$ and imaginary $\Im\omega$ parts for different regions
of the stability diagram shown in figure~\ref{StabDiag}. We chose
a point in each region to show how the dispersion curves change
when moving from one region to another. Each of the subfigures in
figure~$\ref{fig1}$ corresponds to a point with the same name on
the stability diagram. Figure~\ref{fig1}~(a)
demonstrates the behaviour of $\Re\omega$ and $\Im\omega$ in the
region of low partial pressures $\bar{p}_{\text{CO}}$ where the
system is stable. As we see, $\Re\omega < 0$ at any wave number
$k$. In the region of high partial pressures $\bar{p}_{\text{CO}}$,
the dispersion dependences are similar.

In figure \ref{fig1}~(b), we have a realization of the Hopf
bifurcation scenario with $\Im\omega(k_{0})=0.402$,
$\Re\omega(k_{0})=0.002$, where $k_{0}=0$. It is well-known that
the Hopf instability is the local dynamic instability arising in
nonlinear systems with multiple time-scales, and requires the
following conditions: $\Im\omega(k_{0})\neq0$,
$\Re\omega(k_{0})>0$, where $k_{0}=0$ \cite{4}. In the phase space
of the system, it causes a new attractor --- a closed orbit called
the limit cycle \cite{Ebeling}. As a result of the Hopf
bifurcation, an evolution of the system takes place by the states of
the limit cycle. The instability of such type generates patterns periodic
in time.

As figure \ref{fig1}~(c) depicts, depending on the model
parameters, the Turing instability can occur in the system as
well. In contrast to the Hopf bifurcation, the Turing bifurcation is not
dynamic and is caused by the diffusion. It requires
$\Im\omega(k_{\text T})=0$, $\Re\omega(k_{\text T})>0$ where
$k_{\text T}>0$ is the value of wave number~$k$ corresponding to the
second peak of the curve $\Re\omega(k)$ \cite{4}. As is seen in
figure \ref{fig1}~(c), at a certain choice of the diffusion
parameters of the system, namely $\bar{D}_{1}=0.035$,
$\bar{D}_{2}=0.000035$ and $\bar{D}_{3}=1.7$, the condition
$\Im\omega(k_{\text T})=0$, $\Re\omega(k_{\text T})=0.001>0$
becomes true for $k_{\text T}=0.31$. It causes concentration patterns periodic in space
and stationary in time called the Turing
patterns.

In figure \ref{fig1}~(d), dispersion curves are obtained in the
system whose parameters satisfy both the conditions of
the Turing instability and the Hopf one. Thus, the system can lose
stability of the homogeneous state in two ways --- either through
the Hopf bifurcation leading to the temporal patterns formation
(oscillations) in the system or through the Turing bifurcation
that leads to the formation of regular spatial patterns. Moreover,
bifurcations will take place in different ranges of wave numbers
which do not overlap, with $k_{\text{T}}>k_{0}$.

\begin{figure}[!t]
  \centering
a)\raisebox{-.5\height}{\includegraphics[clip=true,keepaspectratio=true,width=0.3\textwidth]{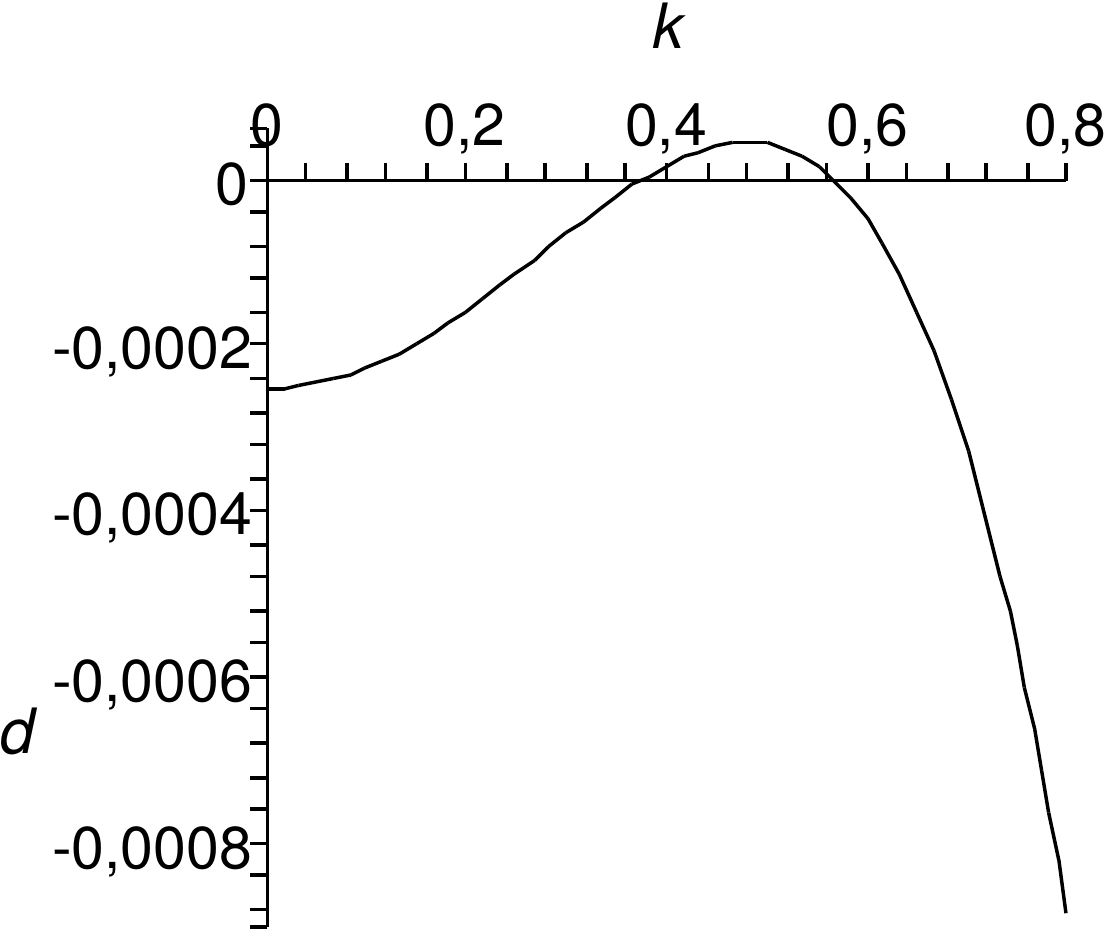}} \qquad\qquad
b)\raisebox{-.5\height}{\includegraphics[clip=true,keepaspectratio=true,width=0.3\textwidth]{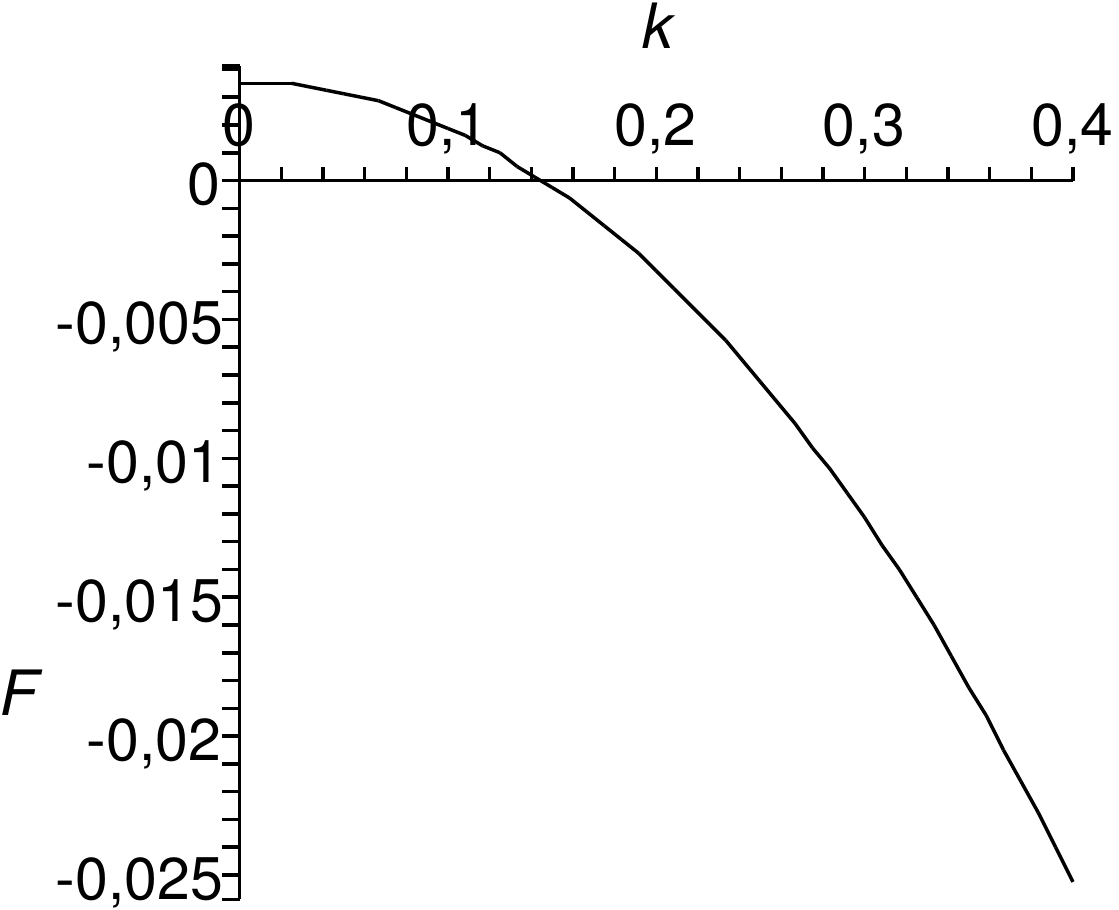}}
  \caption{  Dependences of the coefficients
$d$ (subfigure~a) and $F$ (subfigure~b) on the wave number~$k$ at
a pressure   $\bar{p}_{\rm CO}=0.06$.} \label{d}
\end{figure}

Additionally, using analytical conditions $(\ref{cond})$,
$(\ref{cond2})$, we have plotted $d(k^{2})$ and $F(k^{2})$ as
functions of wave numbers $k$ for a set of the system parameters
corresponding to figure \ref{fig1}~(d). The result is shown in
figure $\ref{d}$~(a) and (b), respectively. We can see that the maxima
of the curves lie in the region of positive values and are as follows:
\begin{align}
d_{\rm max}(k_{\text T}^{2})=5\cdot10^{-5}>0,\qquad F_{\rm
max}(k_{0}^{2})=0.0037>0.
\end{align}
$d_{\rm max}>0$ means the appearance of one real positive
eigenvalue in the system: $\omega_{1}=0.001$, $\omega_{2}=-0.069$,
$\omega_{3}=-0.724$. In the case of $F_{\rm max}>0$, we have two
complex conjugate eigenvalues with a positive real part:
$\omega_{1,2}=0.003\pm 0.419\text{i}$, $\omega_{3}=-0.725$. The
first case corresponds to the Turing bifurcation, and the second
corresponds to the Hopf one.
The numbers used above for the values of model parameters are the
only special cases. The model is more general and, perhaps, it
could be used not only in this particular system, which is
considered as an example. Thus, we have shown that the
Turing  and the Hopf bifurcations can be observed in such systems
despite their non-autocatalyticity.

\subsection{Effect of surface inhomogeneities }

To investigate the effect of inhomogeneities on the surface, we
consider an one-dimensional Pt(110) substrate of a size
$L_{x}=1$~{\textmu}m with various surface phases --- reconstructed
central $1\times2$ phase surrounded by an unreconstructed
$1\times1$ phase. Boundary conditions were chosen assuming that
there is no flow through the boundary of the interval [0,1]. The
initial conditions were set as follows:
\begin{align}
\theta_{\rm CO}(x,t=0)&=\theta_{{\rm CO},\text{s}}\,,\\
\theta_{\rm O}(x,t=0)&=\left\{
\begin{array}{cc}
\theta_{{\rm O},\text{s}}\,, & x<0.3 \quad {\rm and} \quad x>0.7, \\
0, & 0.3<x<0.7,  \\
\end{array}\right.\\
\theta_{1\times1}(x,t=0)&=\left\{
\begin{array}{cc}
1, & x<0.3 \quad {\rm and} \quad x>0.7, \\
0, & 0.3<x<0.7.  \\
\end{array}\right.
\end{align}

Parameters of the reaction and diffusion correspond to the
homogeneous oscillating state. This means that if
the entire surface of the substrate had a uniform structure, the
temporal behavior would be characterized by homogeneous periodic
oscillations of coverages $\theta_{\rm CO}(x,t)=\theta_{\rm
CO}(t)$, $\theta_{\rm O}(x,t)=\theta_{\rm O}(t)$ and by the local
fraction of the surface area in the unreconstructed $1\times1$
phase $\theta_{1\times1}(x,t)=\theta_{1\times1}(t)$.

\begin{figure}[!t]
  \centering
a)\,\,\raisebox{-.5\height}{\includegraphics[clip=true,keepaspectratio=true,width=0.55\textwidth]{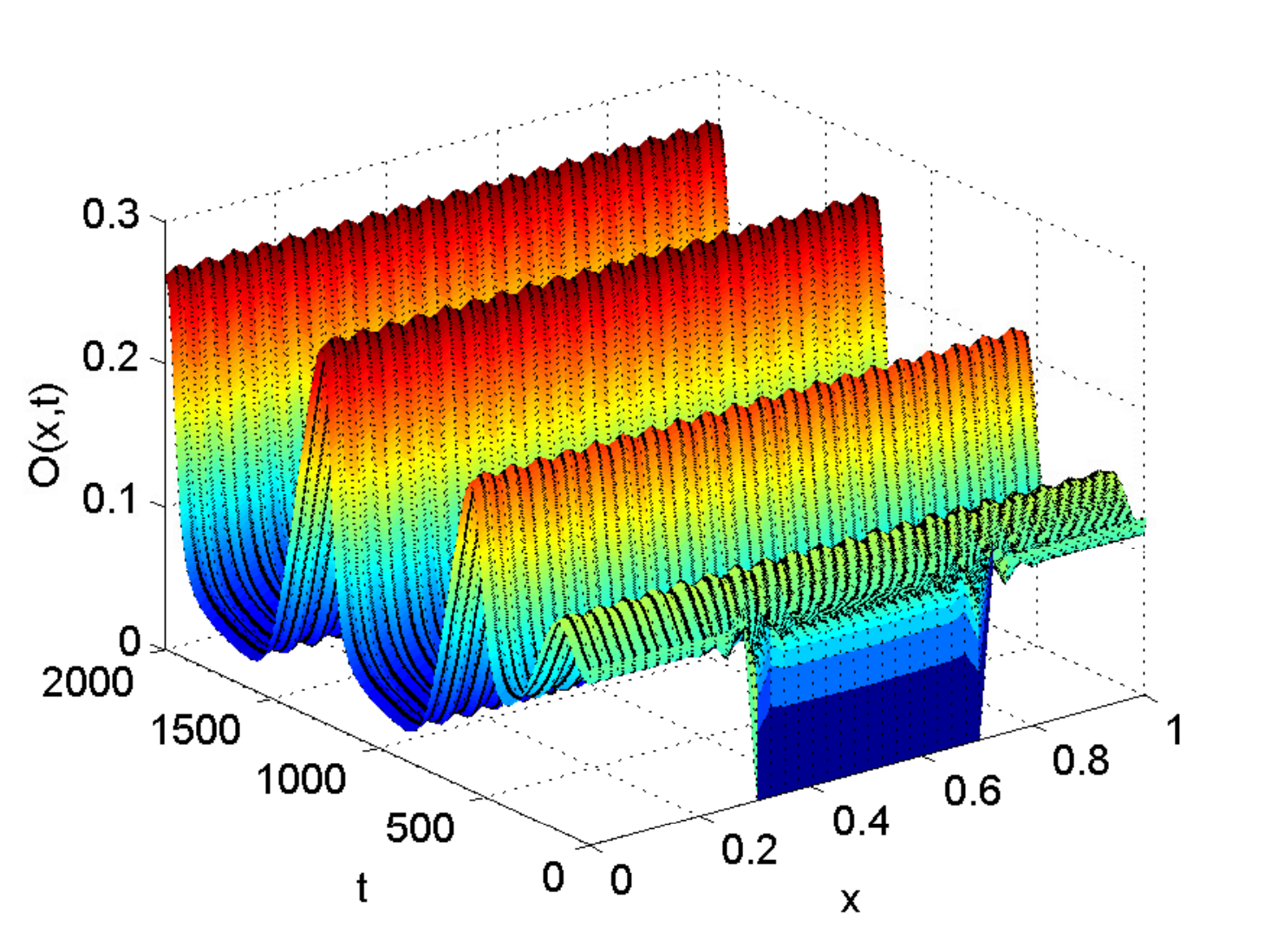}}\\
b)\,\,\raisebox{-.5\height}{\includegraphics[clip=true,keepaspectratio=true,width=0.55\textwidth]{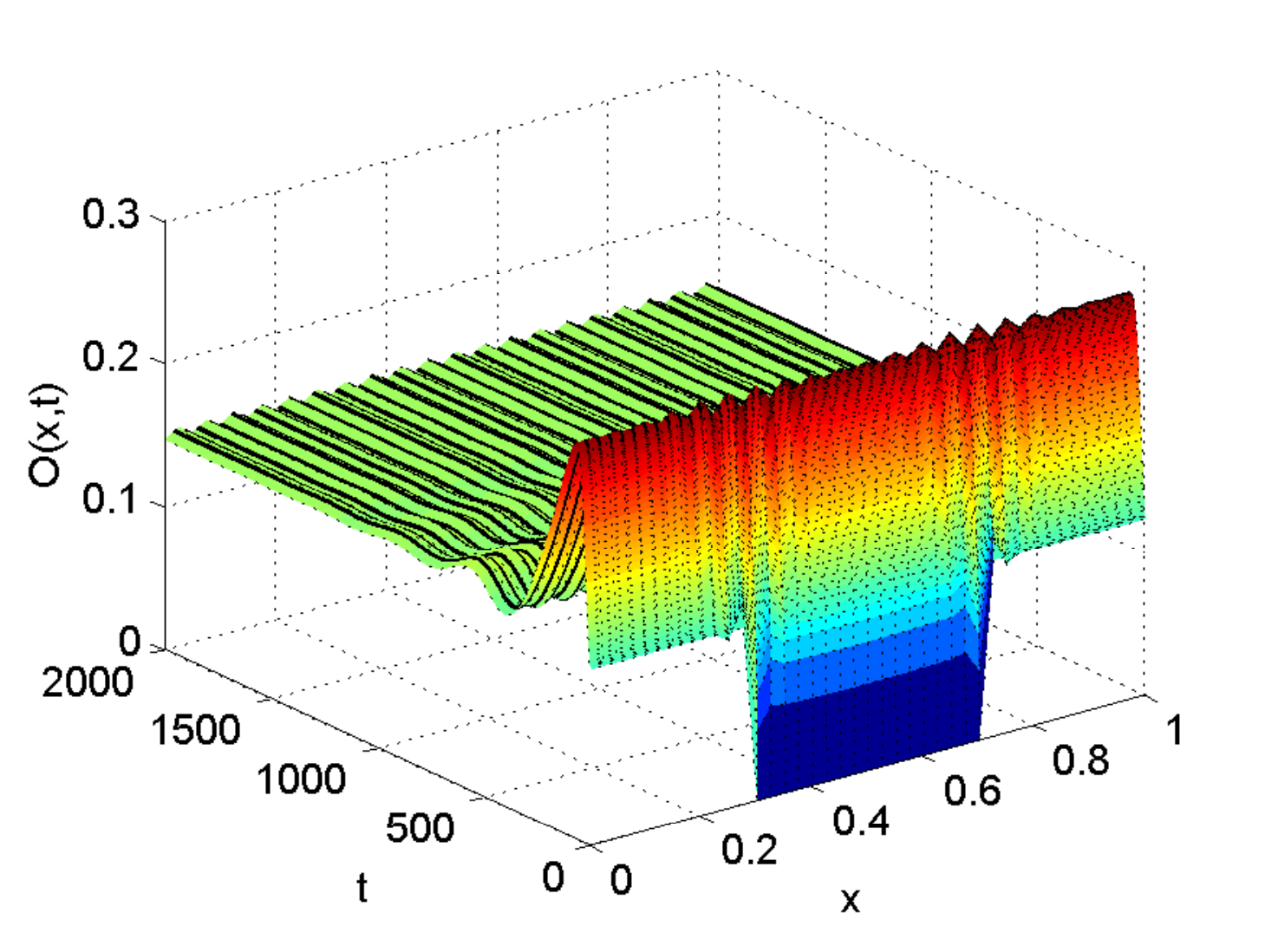}}
  \caption{(Colour online) Spatiotemporal distributions of the oxygen coverage $\theta_{\rm O}$ at pressures $\bar{p}_{\rm CO}=0.06$~(a)
  and $\bar{p}_{\rm CO}=0.053$~(b).}
  \label{fig???}
\end{figure}

For the case of the reconstructed $1\times2$ phase being located
inside the unreconstructed $1\times1$ phase, the gradients of the
adsorbate coverages and of the surface geometry near the
$1\times2$/$1\times1$-interfaces lead to the transition to a
highly nonuniform state, which, in its turn, leads to a deformation of
the wave front. To see this, in figure $\ref{fig???}$ we present
the evolution of the oxygen coverage $\theta_{\rm O}$ at different
values of pressure $\bar{p}_{\rm CO}$.  Figure~\ref{fig???}~(a)
shows that at a partial pressure $\bar{p}_{\rm CO}=0.06$, an
auto-oscillatory regime appears in the system if the
condition~$(\ref{cond2})$ of the existence of the Hopf bifurcation
is satisfied. Figure \ref{fig???}~(b) demonstrates that at
pressure $\bar{p}_{\rm CO}=0.053$, the system evolves to a steady
state through damped oscillations. As we can see, a perturbation of
the initial spatial homogeneous distributions of oxygen coverage
and surface geometry leads to a growth of perturbations by
the Turing mechanism and to the formation of regular
spatiotemporal [figure \ref{fig???}~(a)] and spatial [figure
\ref{fig???}~(b)] patterns with durable coexistence of the regions
with higher and lower oxygen concentrations on the surface.

\begin{figure}[!t]
  \centering
a)\,\,\raisebox{-.5\height}{\includegraphics[clip=true,keepaspectratio=true,width=0.3\textwidth]{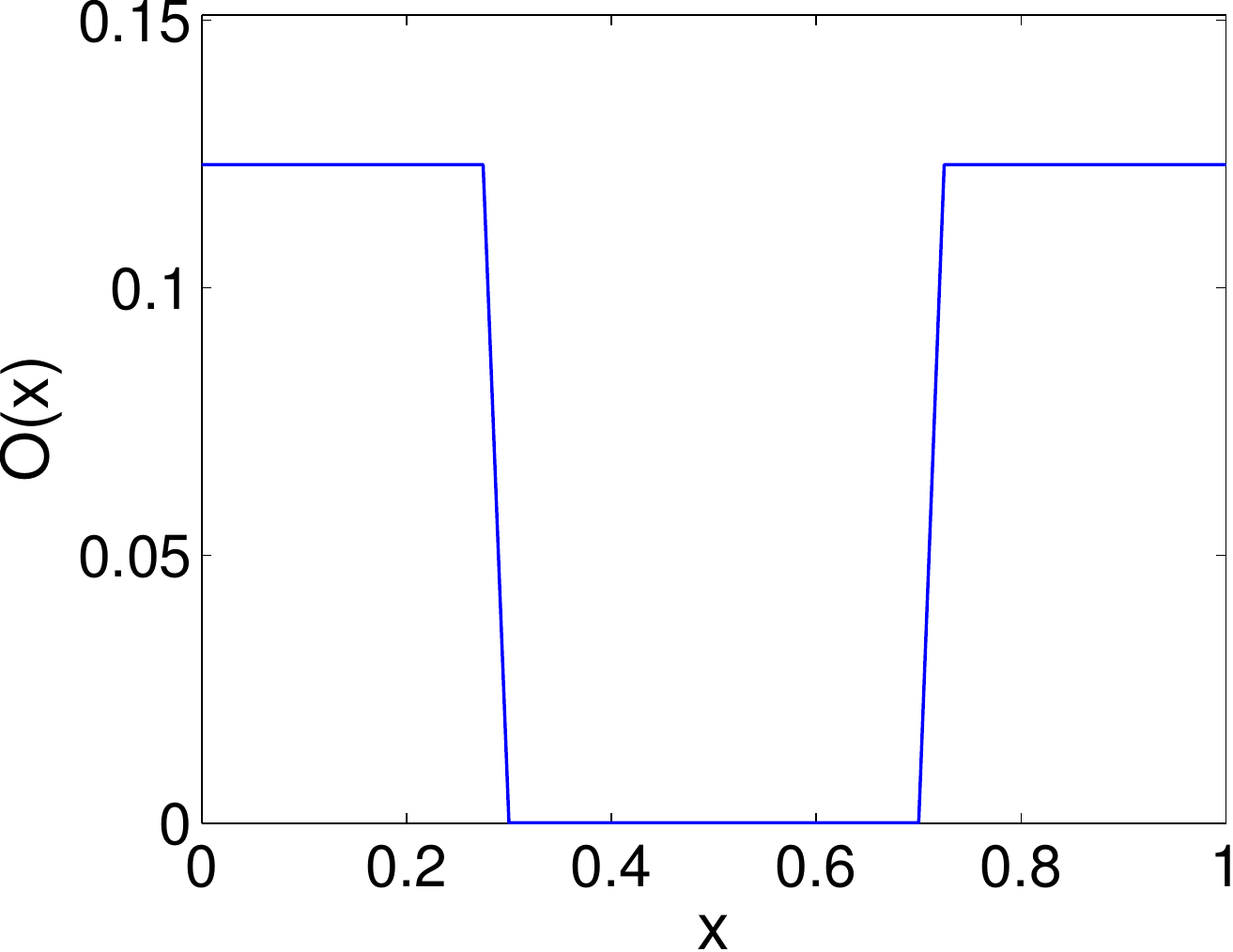}} \qquad
b)\,\,\raisebox{-.5\height}{\includegraphics[clip=true,keepaspectratio=true,width=0.3\textwidth]{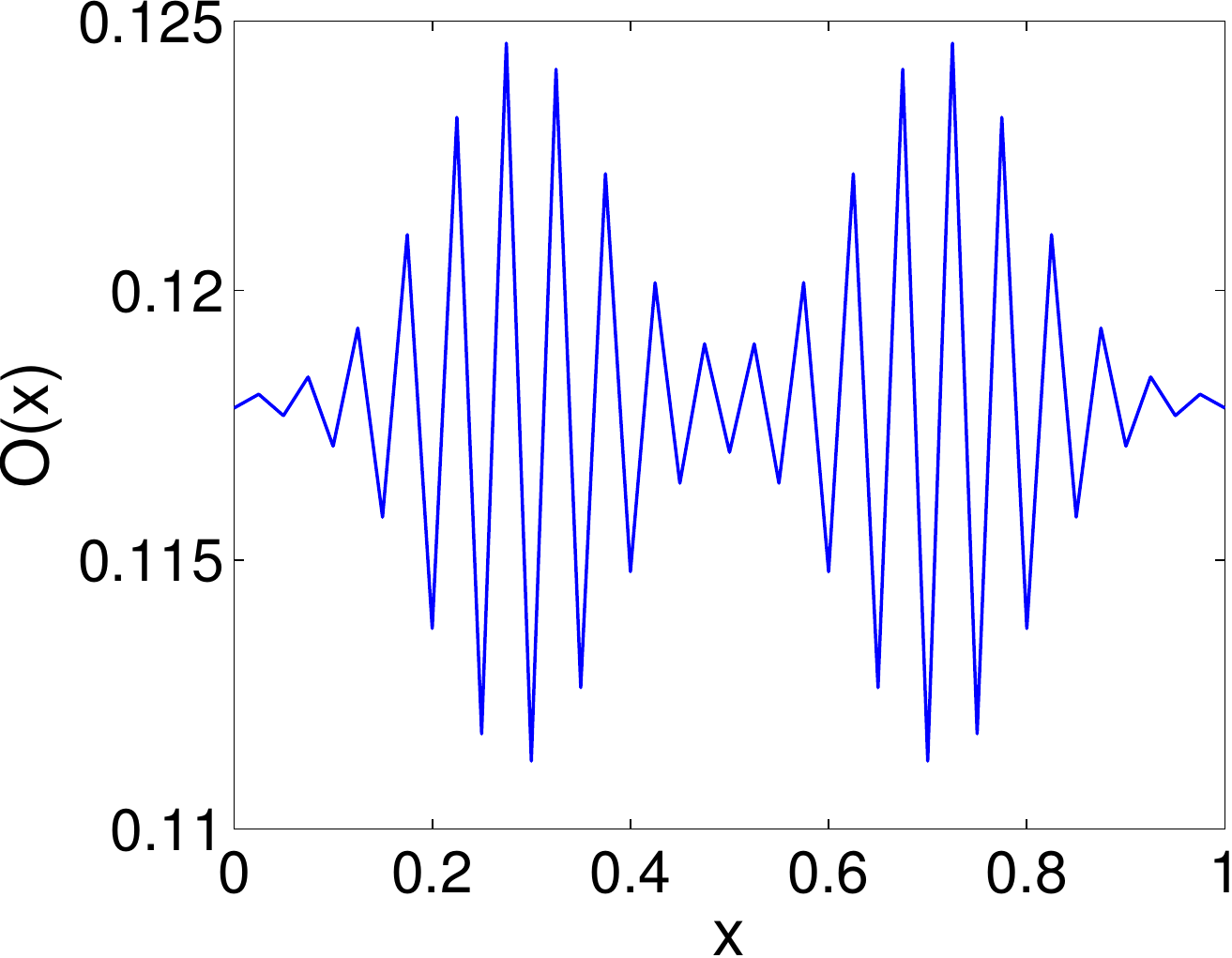}}\vspace{2mm} \\
c)\,\,\raisebox{-.5\height}{\includegraphics[clip=true,keepaspectratio=true,width=0.3\textwidth]{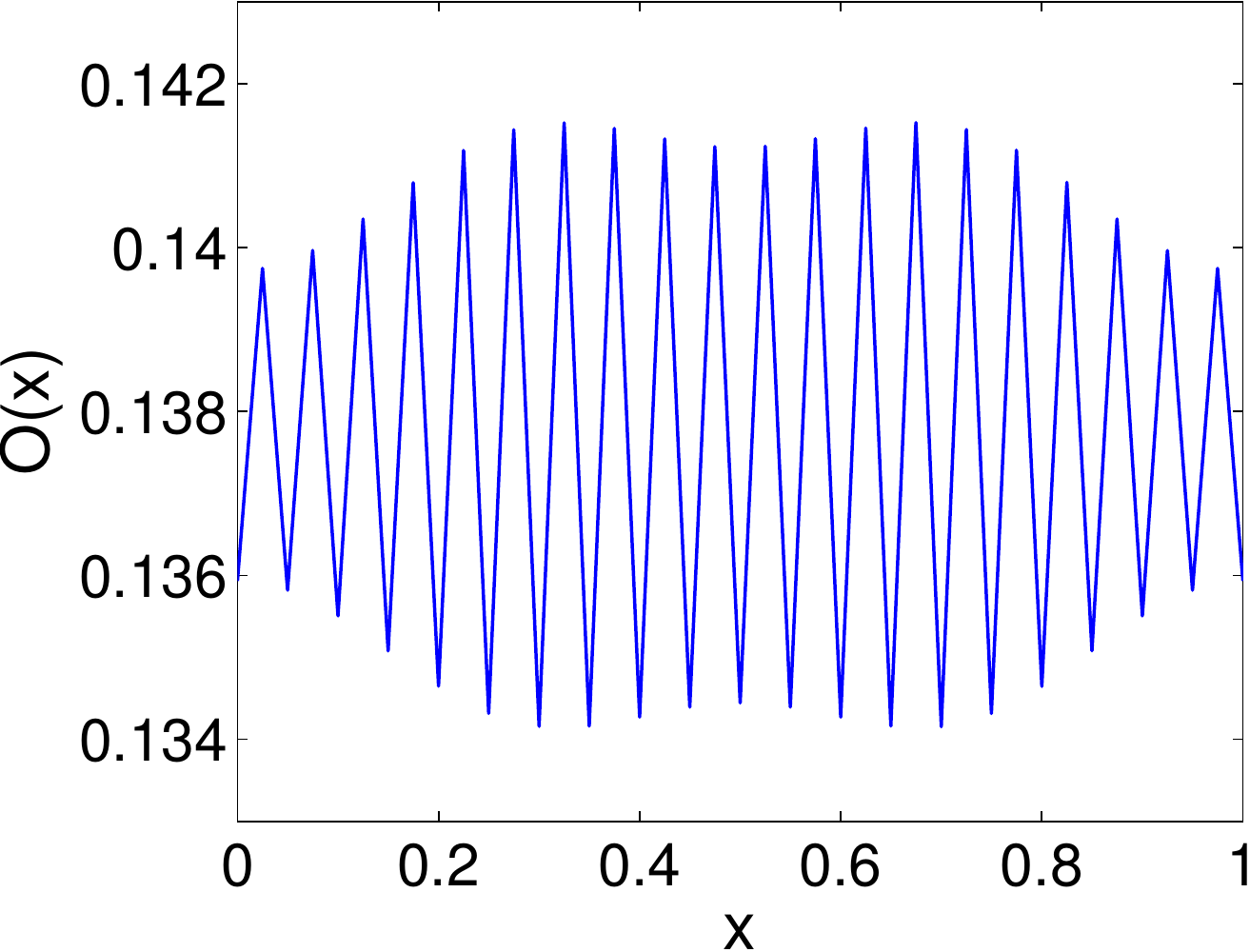}}\qquad
d)\,\,\raisebox{-.5\height}{\includegraphics[clip=true,keepaspectratio=true,width=0.3\textwidth]{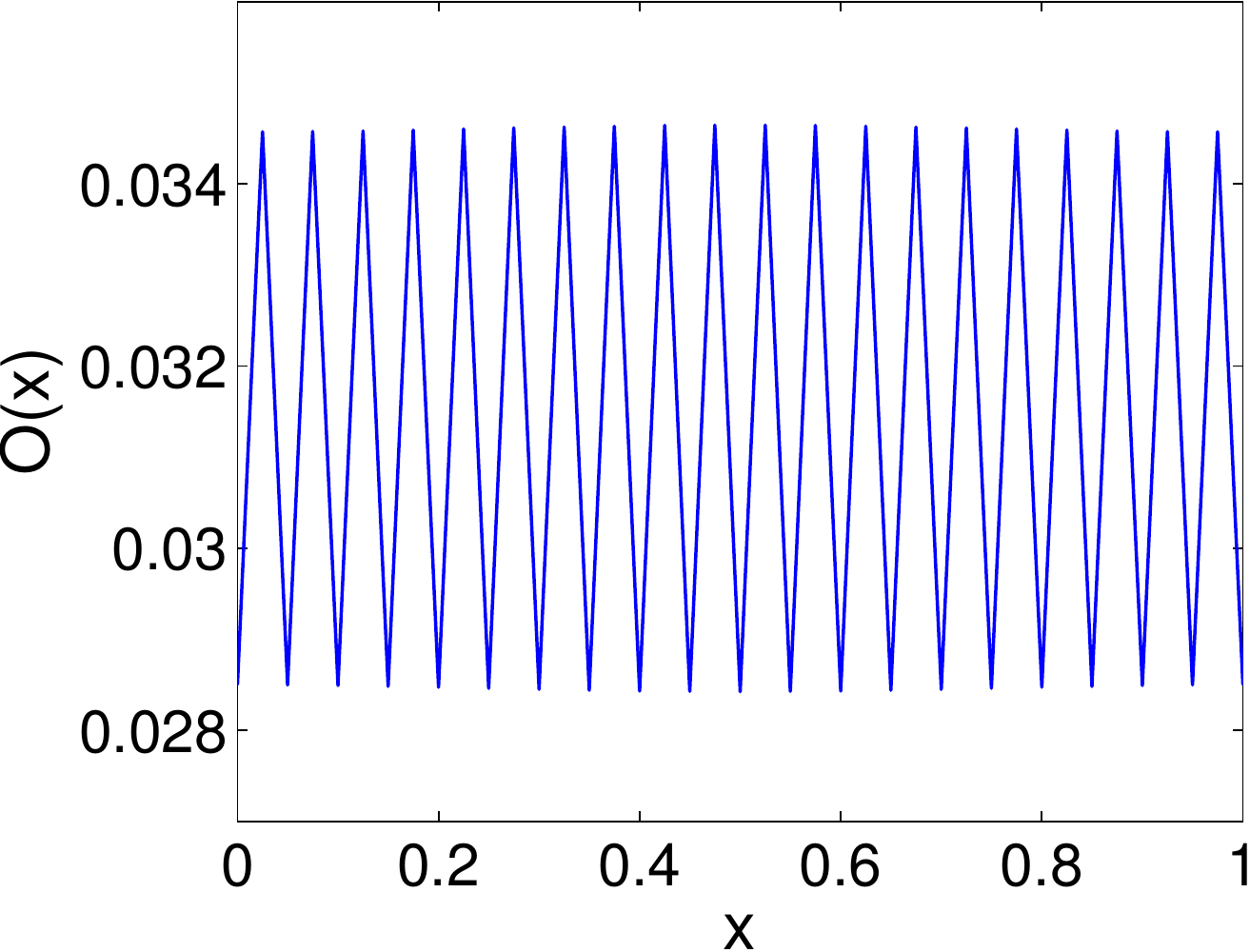}}
  \caption{(Colour online) Time evolution of the spatial distribution of oxygen coverage $\theta_{\rm O}$ at pressure
  $\bar{p}_{\rm CO}=0.06$ at the moments
  $t=0$~(a), $t=50$~(b), $t=200$~(c), and $t=1000$~(d), respectively.}
  \label{fig9}
\end{figure}

\looseness=-1 Figure~$\ref{fig9}$ presents a time evolution of the spatial
distribution of oxygen coverage $\theta_{\rm O}$ at different
moments of time. As is seen in the figure, the distribution of
oxygen on the surface is of an oscillating character. At initial stages
of time evolution, the amplitude of  oscillations is maximal
just at the interfaces. Homogeneous periodic oscillations of
$\theta_{\rm O}$ coverage along the entire surface are observed
over a long period of time.

\begin{figure}[!t]
\centering
\begin{tabular}{ccc}
 &  CO & oxygen \\
\put(0,40){\makebox(0,0){t=750}}&
\includegraphics[clip=true,keepaspectratio=true,width=0.28\textwidth]{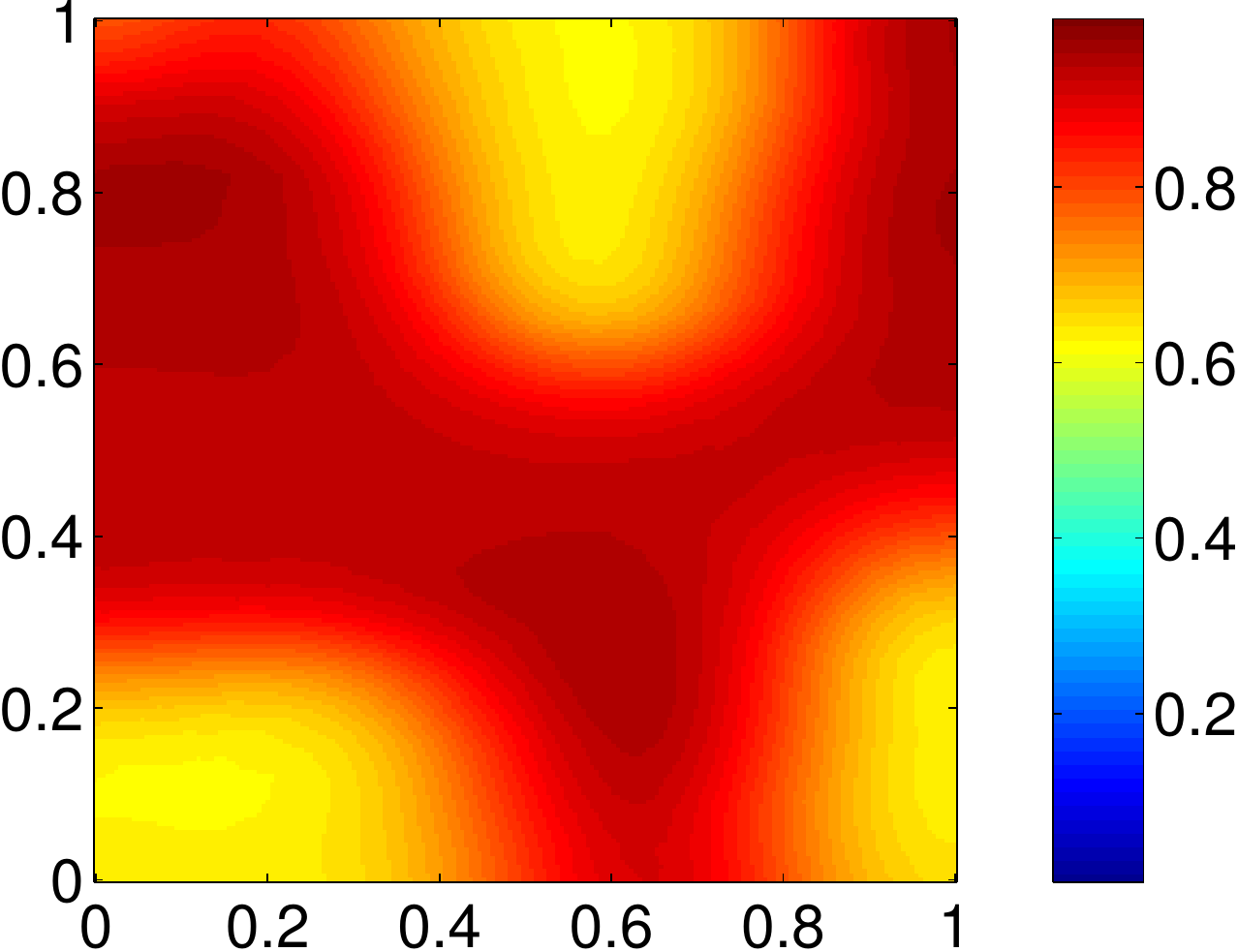}
&
 \includegraphics[clip=true,keepaspectratio=true,width=0.28\textwidth]{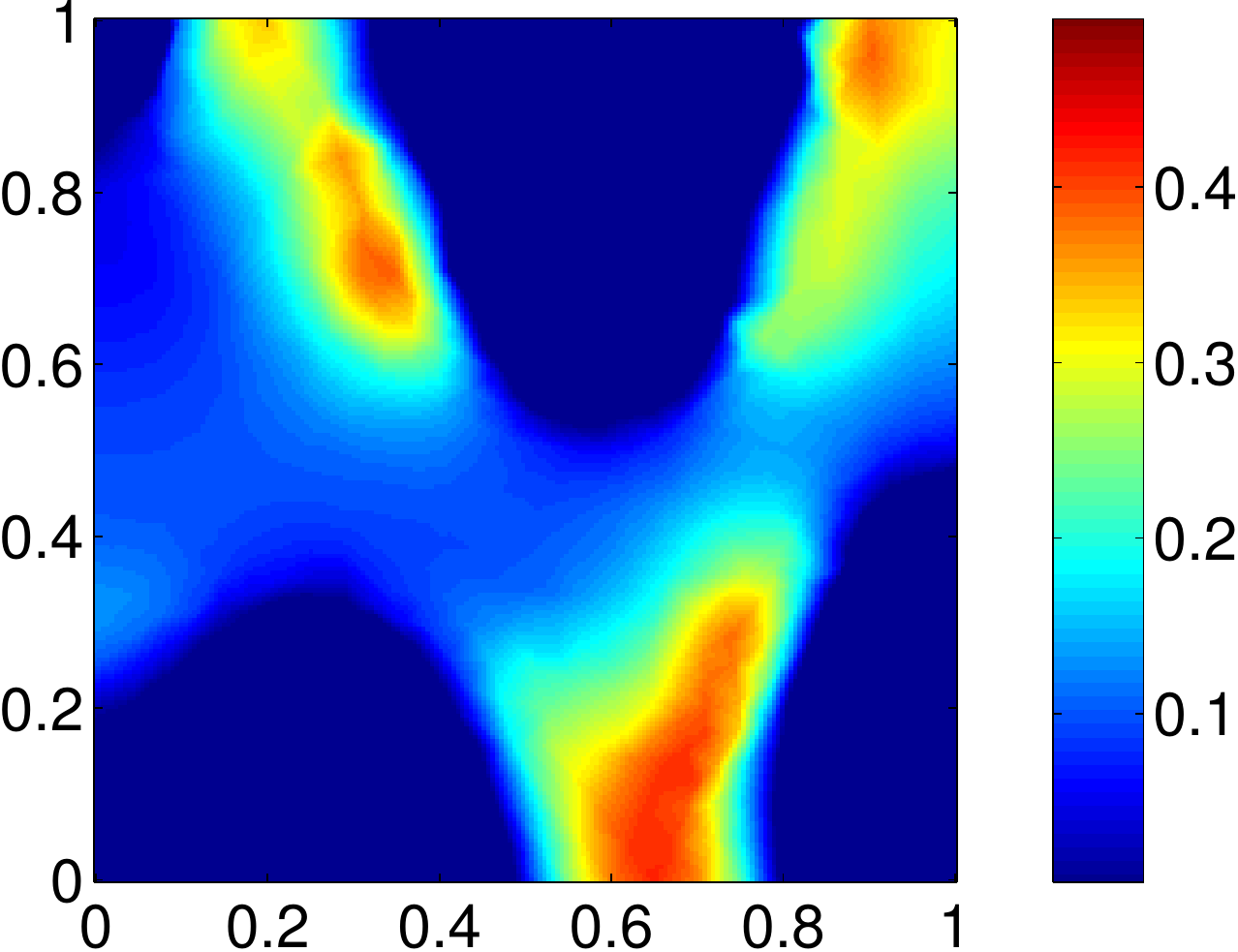}
\\
\put(0,40){\makebox(0,0){t=1000}}&
\includegraphics[clip=true,keepaspectratio=true,width=0.28\textwidth]{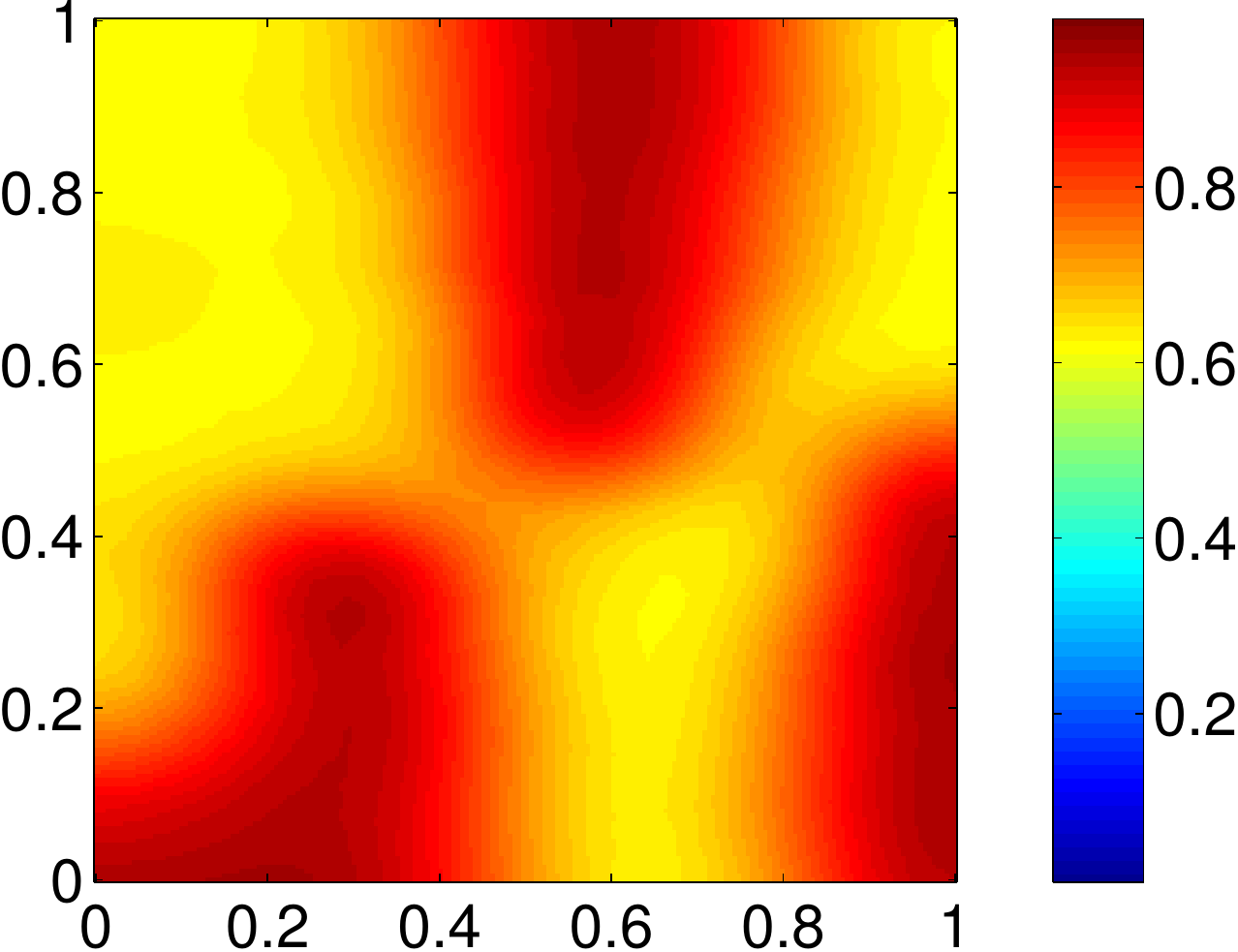}
&
 \includegraphics[clip=true,keepaspectratio=true,width=0.28\textwidth]{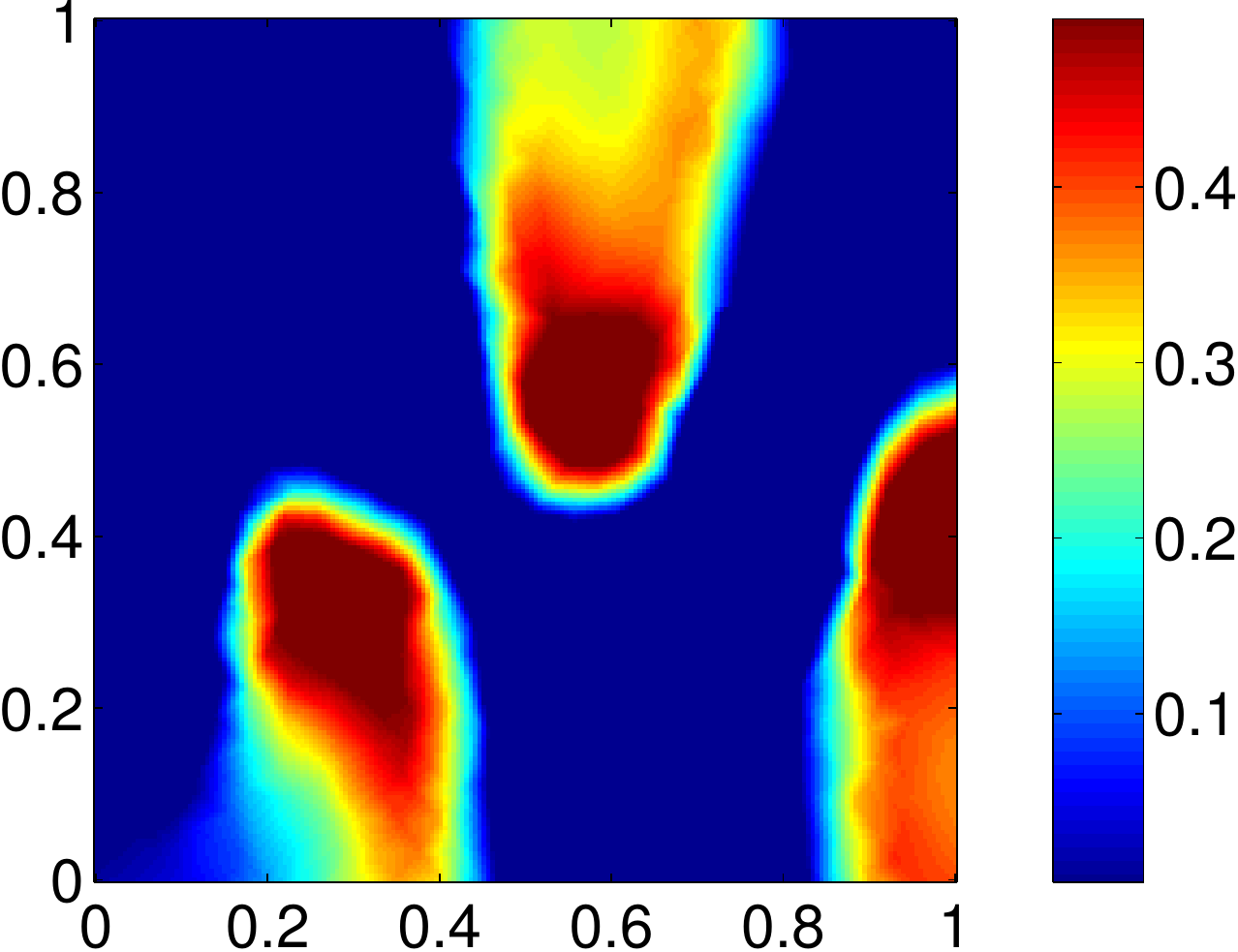}
\\
\put(0,40){\makebox(0,0){t=2500}}&
\includegraphics[clip=true,keepaspectratio=true,width=0.28\textwidth]{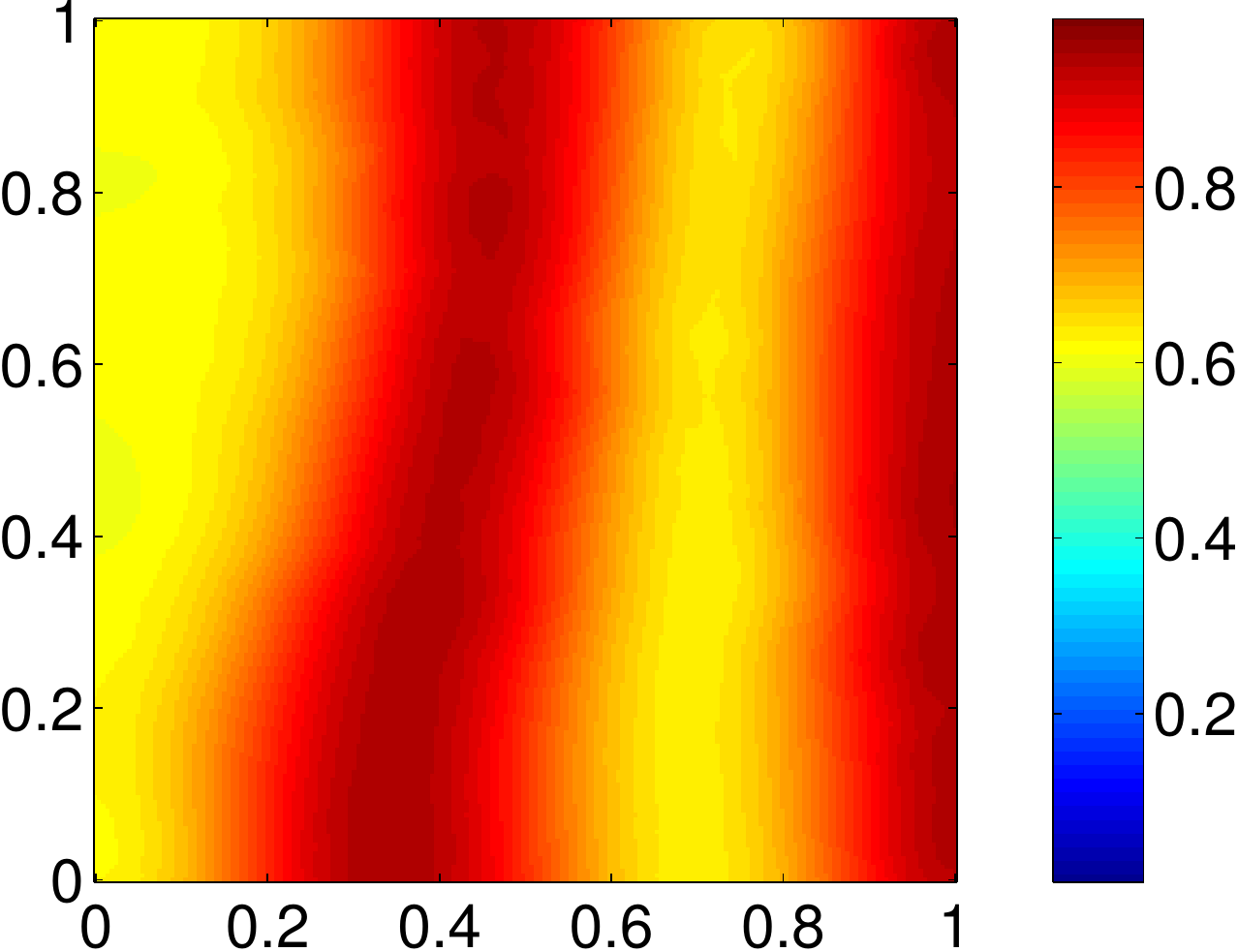}
&
 \includegraphics[clip=true,keepaspectratio=true,width=0.28\textwidth]{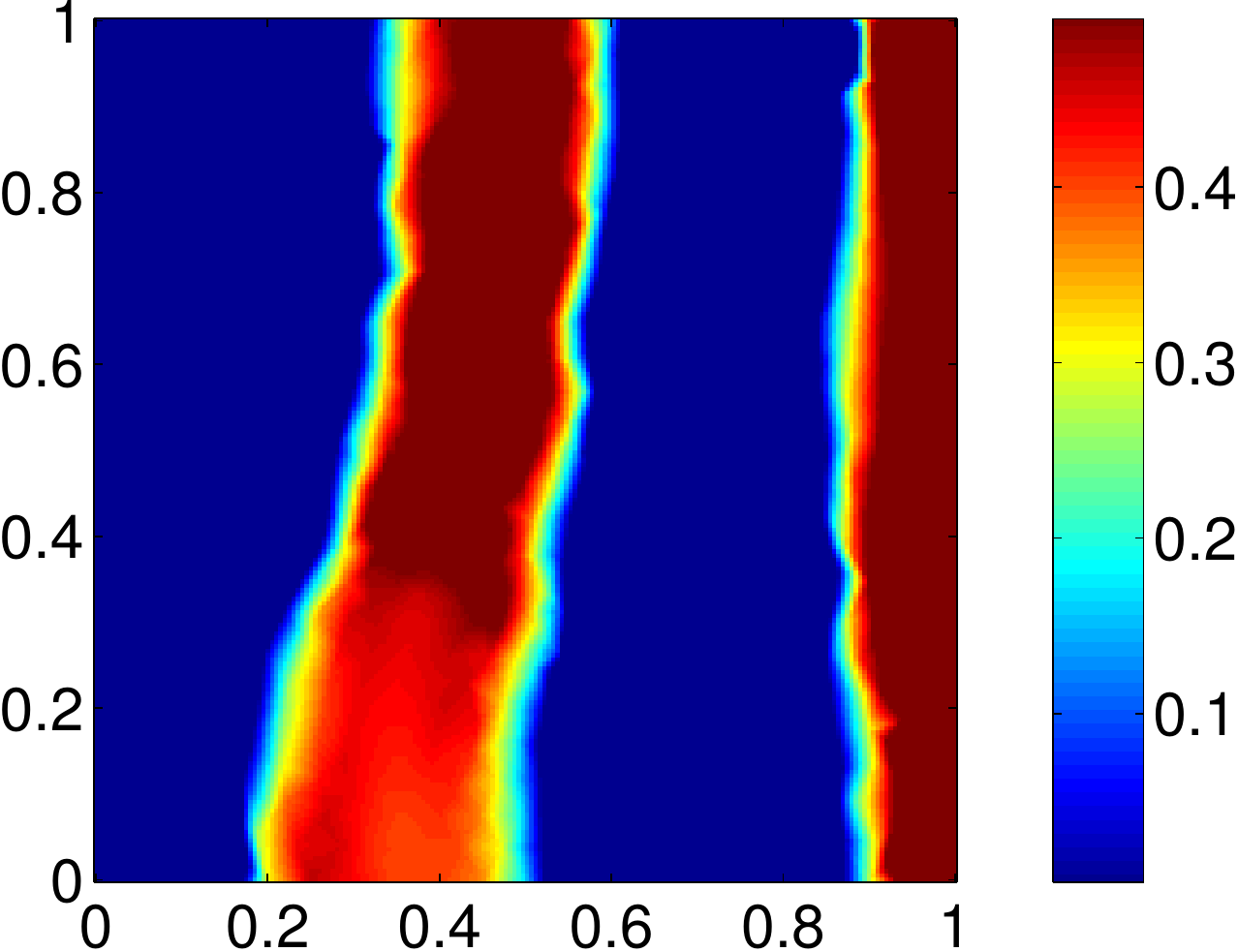}
\\
\put(0,40){\makebox(0,0){t=2750}}&
\includegraphics[clip=true,keepaspectratio=true,width=0.28\textwidth]{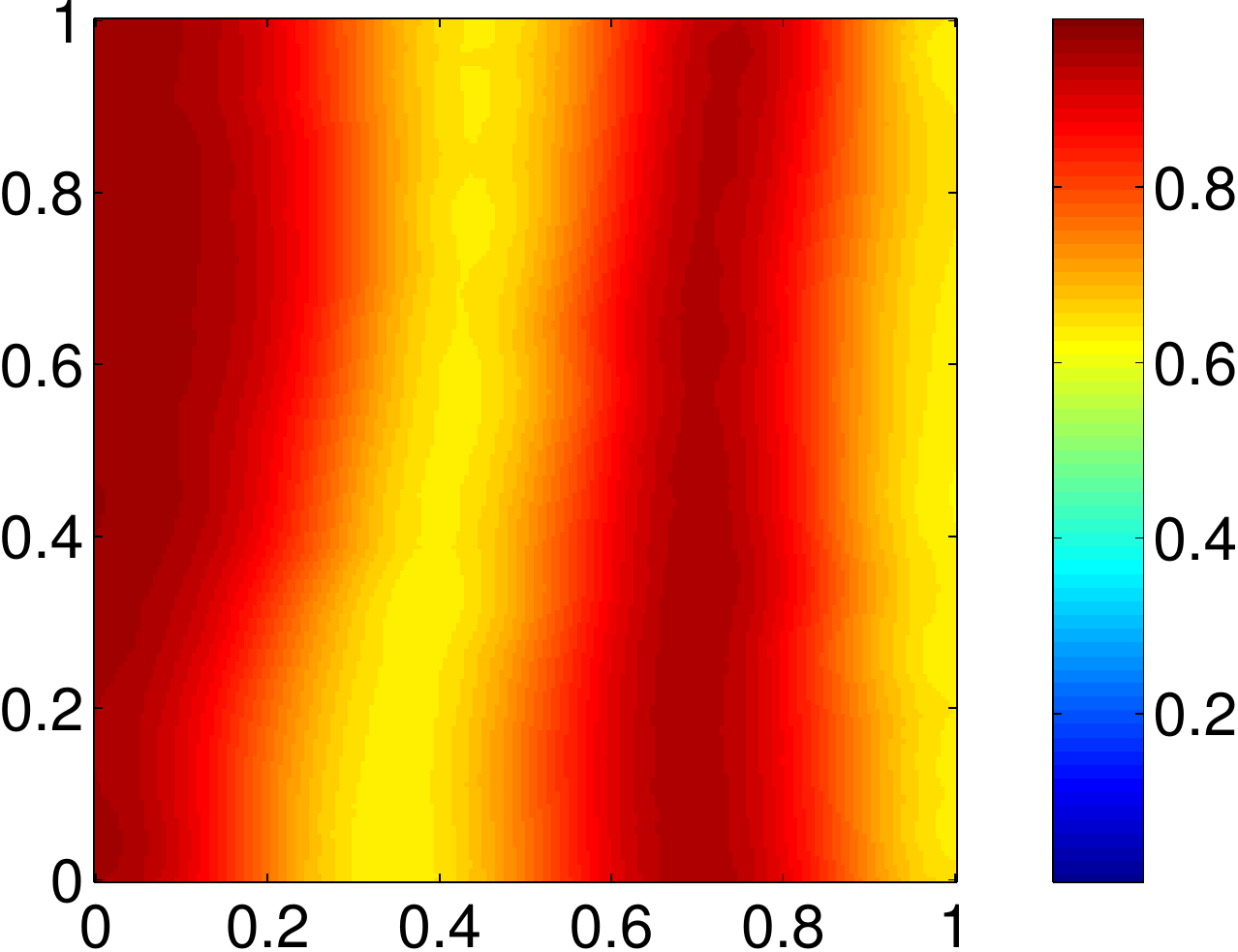}
&
   \includegraphics[clip=true,keepaspectratio=true,width=0.28\textwidth]{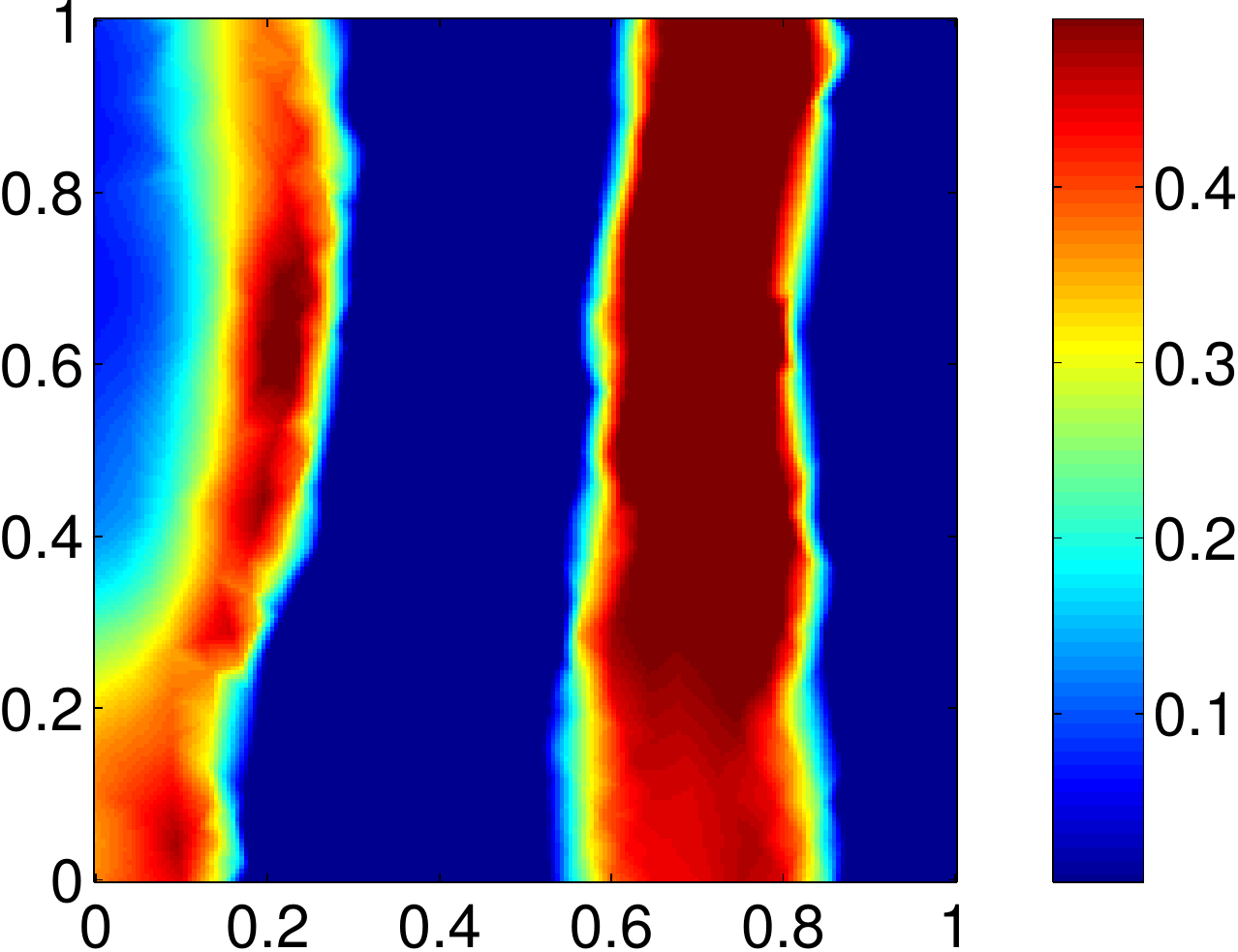}
\\
\end{tabular}
\caption{(Colour online) Examples of surface patterns for $\theta_{\rm CO}$ and $\theta_{\rm O}$
 represented in the form of amplitude maps. The values of pressure $\bar{p}_{\rm CO}=0.06$ and of the coefficient $\bar{D}_{3}=0.3$
   correspond to the region of coexistence of both instabilities. The size of the region is $1\times 1$~{\textmu}m.}
  \label{tab2}
  \end{figure}

Let us consider a case of a two-dimensional Pt(110) surface of
size $L_{x}=L_{y} = 1$~{\textmu}m with different surface phases
--- a reconstructed $1\times2$ phase in the center and an
unreconstructed $1\times1$ phase at the edges. Boundary and initial
conditions were defined similarly to the previous one-dimensional
case.

In the presence of the only Hopf instability [dispersion curves in
figure~\ref{fig1}~(b)], perturbation of the initial spatial
distribution of oxygen and surface geometry leads to the
appearance of surface inhomogeneities at the initial stages of
time evolution. Upon increasing the time, distributions of CO and oxygen become
homogeneous in space but oscillating in time
--- we observe homogeneous periodic time oscillations of both
coverages along the entire surface. Distribution of the surface
phase $\theta_{1\times1}$ becomes homogeneous in space and
stationary in time.

In the case of the coexistence of both instabilities [dispersion
curves in figure~\ref{fig1}~(d)], we observe the emergence of
spatiotemporal patterns for the $\theta_{\rm CO}$ and $\theta_{\rm
O}$ coverages. The distribution of the surface phase
$\theta_{1\times1}$ becomes, as in the previous case, almost
homogeneous in space and stationary in time. In
figure~$\ref{tab2}$, several patterns are presented in the form of
amplitude maps obtained at different moments of time  using
numerical simulations. 
We observe oscillating nonuniform
distributions of adsorbate coverages with  coexistence of the
regions of high and low concentrations  on the surface.  These are traveling waves of chemical concentrations observed during experimental investigations  of the catalytic CO oxidation
under low pressure conditions on Pt(110), see, e.g. \cite{exp1,exp2}. Upon a subsequent increase of time $ t> 2750 $, no fundamentally new spatial distributions of CO and oxygen were observed. In general, the appearing
spatiotemporal patterns  depend on a specific choice of
the system parameters.

\section{Conclusions}
\label{Concl}

The catalytic carbon monoxide oxidation reaction model taking
diffusion processes on the Pt(110) surface into account has been
considered. The dispersion dependences $\Re\omega$ and $\Im\omega$
on the wave number~$k$ have been built. Despite  the CO
oxidation reaction being non-autocatalytic, we have shown that the
analytic conditions of the existence of the Turing and the Hopf
bifurcations can be satisfied at certain values of the system
parameters. Thus, the system may lose its stability in two ways
--- either through the Hopf bifurcation leading to the formation
of temporal patterns in the system or through the Turing
bifurcation leading to the formation of regular spatial patterns.
The regions corresponding to the existence of a particular bifurcation
were identified in the parametric space. At a simultaneous
implementation of both scenarios, spatiotemporal patterns for CO
and oxygen coverages have been observed in the system. The
emergence of these instabilities are associated with an
interaction of surface phase transitions and diffusion processes
that spatially couple the system. The appearing spatiotemporal patterns 
 depend on a specific choice of the system parameters. In
general, the case of both instabilities being possible is rather
complicated and deserves a special attention.

\ukrainianpart

\title{Формування просторово-часових структур у трикомпонентній моделі реакції окислення
монооксиду вуглецю}
\author{I.С.~Бзовська, I.M.~Мриглод }
\address{Інститут фізики конденсованих систем НАН України, вул. Свєнціцького, 1, 79011 Львів, Україна}

\makeukrtitle

\begin{abstract}
\tolerance=3000%
Досліджуються механізми формування просторово-часових структур у
каталітичній реакції окислення CO з урахуванням процесів дифузії
на неоднорідній поверхні Pt(110), яка містить структурно відмінні
ділянки, що утворюються під час СО-індукованого переходу від
реконструйованої $1\times 2$ фази до об'ємної $1\times1$ фази.
Незважаючи на те, що реакція окислення СО не є автокаталітичною,
ми показали, що аналітичні умови існування біфуркацій Тюринга та
Хопфа виконуються в таких системах. Тобто, система може втрачати
стійкість двома шляхами: або через біфуркацію Хопфа, що веде до
утворення в системі часових структур, або через біфуркацію
Тюринга, що призводить до формування регулярних просторових
струк\-тур. При одночасній реалізації обох сценаріїв у системі
отримано просторово-часові структури для величин покриття СО та
кисню.

\keywords реакційно-дифузійна модель, просторово-часові структури,
біфуркація Хопфа, біфуркація Тюринга

\end{abstract}

\end{document}